\documentclass{aa} 

\usepackage{graphicx}
\usepackage{txfonts}
\usepackage{mhchem}
\usepackage{bm}
\usepackage{comment}
\usepackage[small,bf]{caption}
\usepackage[subrefformat=parens]{subcaption}
\begin{document}

   \title{Monte Carlo Simulation of Sugar Synthesis 
   on Icy Dust Particles Intermittently Irradiated by UV in a Protoplanetary Disk}
   \author{Hitoshi Takehara
          \inst{1,2}
          \and
          Daigo Shoji
          \inst{3}
          \and
          Shigeru Ida
          \inst{1}
           }
   \institute{Earth-Life Science Institute, Tokyo Institute of Technology, Meguro-ku, Tokyo 152-8550, Japan 
         \and
         Department of Earth and Planetary Sciences, Tokyo Institute of Technology, Meguro-ku, Tokyo 152-8551, Japan 
         \and
         Institute of Space and Astronautical Science, Japan Aerospace Exploration Agency, Chuo-ku, Sagamihara, Kanagawa 252-5210, Japan
           }
   \date{DRAFT:  \today}
  \abstract{   
While synthesis of organic molecules in molecular clouds or protoplanetary disks is 
complex, observations of interstellar grains, analyses of carbonaceous chondrites, and UV photochemistry experiments 
are rapidly developing and providing constraints on and clues to the complex organic molecule synthesis in space. 
It motivates us to construct a theoretical synthesis model.
} 
{We develop a new code to simulate global reaction sequences of organic molecules
to apply it for sugar synthesis by intermittent UV irradiation on the surface of
icy particles in a protoplanetary disk.
Here we show the first results of our new simulation.
}
{
We apply a Monte Carlo method to select reaction sequences from all possible reactions, 
using the graph-theoretic matrix model for chemical reactions and modeling
reactions on the icy particles during UV irradiation.
}
 {
We here obtain the results consistent with the organic molecules in carbonaceous chondrites and obtained by the experiments,
however, through a different pathway from 
the conventional formose reactions previously suggested.
During UV irradiation, loosely-bonded O-rich large molecules are continuously created and destroyed.
After UV irradiation is turned off, the ribose abundance rapidly increases,
through the decomposition of the large molecules with break-ups of O-O bonds and 
replacements of C-OH by C-H to reach O/C = 1 for sugars.
The sugar abundance is regulated mostly by the total atomic ratio H/O of starting materials,
but not by their specific molecule forms. Deoxyribose is simultaneously synthesized,
and most of the molecules end up with complex C-rich molecules.
 }
{}
     \titlerunning{Sugar Synthesis on UV-irradiated Icy Dust Particles}
   \maketitle

\section{Introduction} \label{sec:intro}

In carbonaceous chondrites, polyhydroxylated compounds, and sugar derivatives,
including sugar alcohols (reduced sugars) or sugar acids (oxidized sugars),
were detected \citep{cooper2001, cooper2016}, in addition to amino acids and nucleobases
\citep{Martins2008,Martins2015}.
Recently the detection of ribose (which is 
a sugar C$_k$(H$_2$O)$_k$ with $k=5$; 5-C sugars) in meteorites was also reported \citep{furukawa_2019}.
According to the findings, UV photochemistry
experiments \citep{meinert,nuevo} 
have been done to explore a possibility of sugar synthesis on the surface of icy dust particles 
in the interstellar or protoplanetary disk environments.
They successfully produced
ribose, deoxyribose 
and related sugars in the UV irradiated products formed from simple molecules.
These findings suggest a possible hypothesis that these molecules were synthesized in molecular clouds or
protoplanetary disks and delivered to the Earth
(\citealp{oro_1961, chyba_1992}).

\begin{figure}[ht]
\begin{center}
\fbox{
\includegraphics[width=0.9\linewidth]{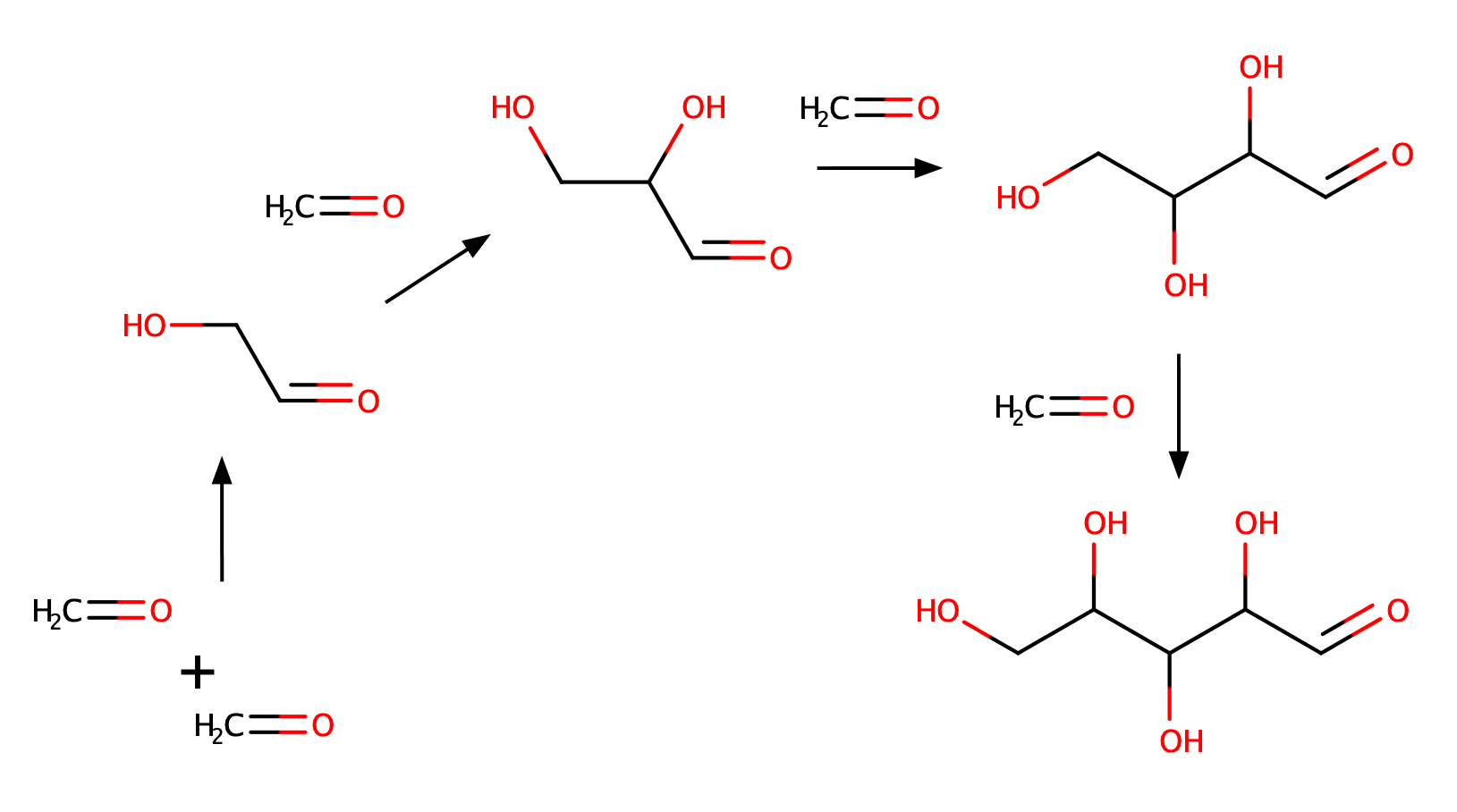}
}
\caption{Schematic diagram of typical formose-type reactions.}
\label{formose_image}
\end{center}
\end{figure}

\citet{meinert} performed a photochemistry experiment by UV with 10~eV on
a plate initially composed of $\mathrm{CH_3OH}$, $\mathrm{H_2O}$ and $\mathrm{NH_3}$
at 78 K which mimics the interstellar/protoplanetary icy dust surface environments,
and reported that many kinds of sugars were synthesized, including ribose, in the final products
at room temperature after UV is turned off. 
The authors suggested that the diverse sugar molecules are formed by formose-type reactions (Fig.~\ref{formose_image}).
Formose-type reactions are well known as the bottom-up reaction for sugar synthesis by aldol condensations to produce various molecules with linear and branched structures. 
However, there is no clear evidence in their experiment that sugars 
are formed mostly from formose-type reactions.
Another similar experiment on sugar synthesis (except at 12~K instead of at 78~K) reported 
the detection of deoxysugars in the UV photochemistry products \citep{nuevo}.
In general, deoxysugars are not synthesized via formose-type reactions \citep{butlerow1861, breslow1959}.
Since deoxysugar derivatives are also detected in meteorites \citep{cooper2001, nuevo}, a different reaction from formose-type reactions may also contribute to the synthesis of sugar molecules on icy dust particles.


\begin{figure}[ht]
\begin{center}
\includegraphics[width=9cm]{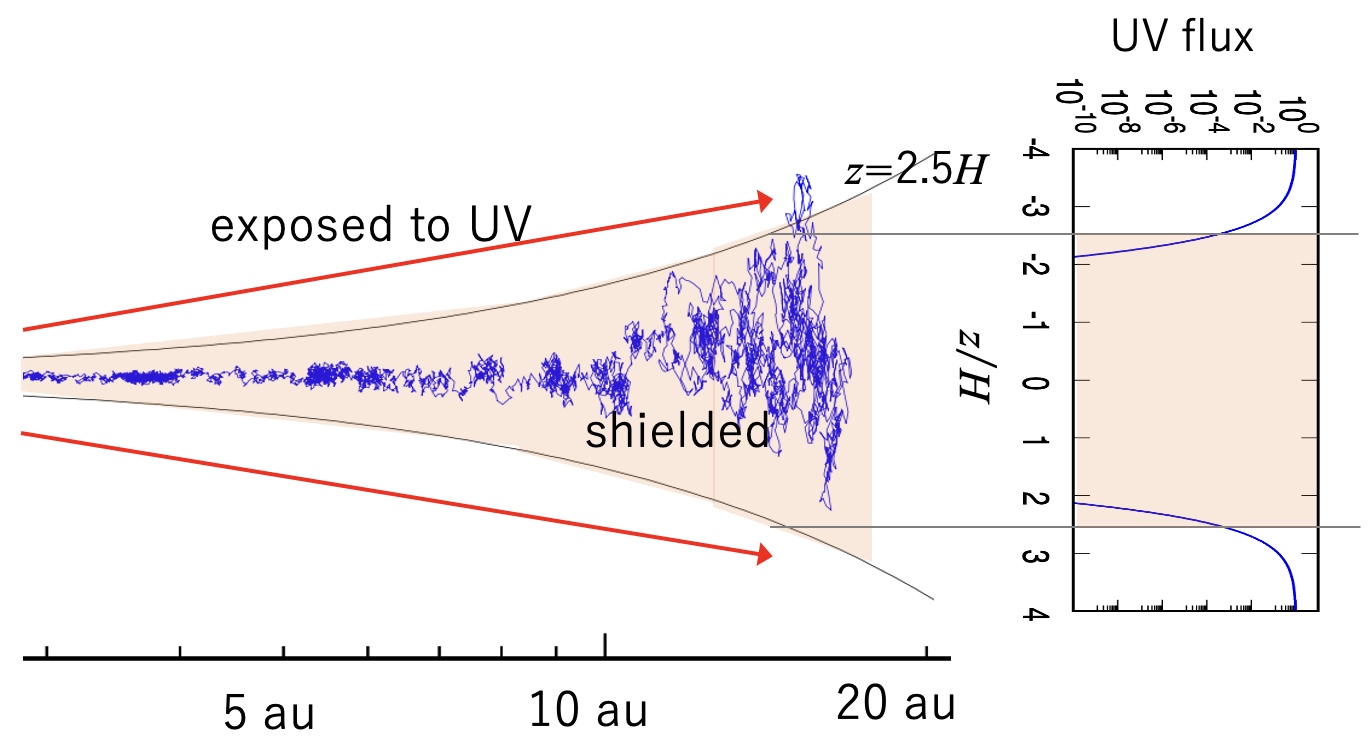}
\caption{A schematic illustration of a trajectory of an icy particle
in a turbulent protoplanetary disk.
The trajectory is represented by the zigzagged blue line.
The shaded region represents a typical UV-shielded vertical height $|z| < 2.5 H$,
where $H$ is the disk scale height.
The right panel shows a typical vertical ($z$) attenuation of UV from a central star by the protoplanetary disk gas. 
The particle trajectory is simulated by the method of
\citet{Okamoto2022} with the effect of particle growth based on \citet{Sato2016}.  
The particle is released at 15 au on the mid-plane
and its orbit evolves by diffusion induced by the disk turbulence and
inward advection due to gas drag. 
}
\label{fig:disk}
\end{center}
\end{figure}

In this paper, 
we focus on synthesis pathways of sugars or related molecules on the UV-irradiated surface of icy particles
in relatively warm temperature ($\sim$ 50--100~K) environments such as in a protoplanetary disk.
Particles occasionally diffuse to a highly upper layer of the disk to receive UV radiation from the host star,
while for most of the time, the particles stay in the disk regions where UV radiation is shielded \citep[][also see Fig.~\ref{fig:disk}]{ciesla_2012}.
Each UV irradiation experiment would correspond to a full cycle of travels of the icy particle out of and 
back to the UV shielded region; UV photon flux for the disk environments may be up to $10^2$ times higher than that for diffuse interstellar clouds
and up to $10^5$ times higher than that for cold dense interstellar clouds \citep{ciesla_2012,nuevo}.
In these UV irradiation experiments, the photochemical products were
heated up to room temperature.
It is not clear when the detected sugar molecules were synthesized, during UV irradiation or after that at room temperature.

We will show in this paper that chemical reactions of UV photochemical products 
significantly proceed after the UV irradiation is turned off.
We will also show the products do not substantially depend on temperature after UV irradiation, as long as the temperature is in a rather broad range of 
$\sim 10$--1000~K.

In order to identify the reaction pathways to form complex organic molecules such as sugars, 
we employ computational simulations.
In the computational studies of the synthesis of complex organic molecules 
on the surface of interstellar icy particles, 
the rate equation method has been often used \citep[e.g.,][]{chang_2016, garrod_2019, jin_2020}.
In these studies, setting target molecules (for example, methyl formate $\mathrm{CH_3OCHO}$, 
dimethyl ether $\mathrm{CH_3OCH_3}$, acetaldehyde $\mathrm{CH_3CHO}$, and methanol $\mathrm{CH_3OH}$), 
a reaction network is set up based on the data of the reaction rate coefficients.
Because cold environments, where most molecules are not sufficiently mobile, are often considered, 
a chemical kinetic model on the dust surface is sometimes combined with the reaction network model
\citep{chang_2016}.
In the kinetic model, the surface and the bulk of icy particles are divided into sites and 
Monte Carlo calculation is applied for molecule hopping from one site to another
using an Arrhenius-type
exponential probability with the hopping barrier potential energy.

Because we consider relatively warm environments with $T \sim$~50--100~K, 
we do not adopt such spatially limited kinetics
and for simplicity we assume that all the species 
can interact with one another.  
Instead, we apply the Monte Carlo approach for chemical reactions
without preparing a reaction network.
To construct the Monte Carlo chemical-reaction scheme, 
we utilize a graph-theoretic matrix framework to represent molecules and chemical reactions that was originally developed by \citet{du} (Sect.~\ref{subsec:matrix}).
The matrix model has been used to find pathways to synthesize target products
often for industrial purposes in the past studies \citep[e.g.,][]{du,Habershon_2015,Habershon_2016,kim_2018,ismail}.
The method used in \citet{du} is ``backward," in which reactants were investigated from target molecules. \citet{Habershon_2015,Habershon_2016} and \citet{ismail} considered ``forward" path (from reactants to targets). These works evaluate each synthesis referencing basal chemical reactions confirmed in laboratory experiments.

Although the database that registers reactions in laboratory experiments is very useful to assess synthesis paths, our primary purpose is to investigate how sugars (in particular, ribose) are synthesized in a broad range of conditions (environments) from the protoplanetary disk to the early Earth. 
Thus, in this work, we try to evaluate chemical reactions without the reaction database. 
We do not restrict our study to only sugar synthesis, but also
the synthesis of photochemical products in general including 
sugar alcohols, deoxysugars, and even Insoluble Organic Matter (IOM) 
that have been found in carbonaceous chondrites.
We estimate the relative abundance
of sugars, sugar alcohols, and deoxysugars
and study in a general manner.
In order to achieve this, we need a global survey in parameter space,
so that we use the method without using the pre-registered reaction database.
In this sense, our approach is a forward-type model and 
we perform Monte Carlo calculations with  
an Arrhenius-type exponential weighting (Sect.~\ref{weighting}). 

Another forward-type method without a database is quantum chemistry calculation. 
For Miller-type amino acids synthesis \citep{miller1953, miller1955},
the quantum chemistry simulations were applied \citep{wang_ab_initio,saitta_2014}.
Quantum chemistry calculations can evaluate synthesis paths of molecules accurately, and they have been widely used in the field of chemistry.
Because reactions occur freely without preordained reaction coordinates or elementary steps, 
new reaction pathways can be discovered only by setting initial molecules and simulation environments.
\citet{wang_ab_initio} used an \textit{ab initio} nanoreactor with repeated high pressure 
by a virtual piston that is 
equivalent to $\sim \! \! 10^4$ K and found a new pathway to form glycine.
\citet{saitta_2014} performed the calculation at 400 K and 
argued that electric field accompanied with the electric discharge
is a key factor for glycine formation. 
In the work with the graph-theoretic method, \citet{kim_2018} also used quantum chemistry calculation for kinetic analysis of each reaction.

Quantum chemistry calculation is a commonly-used powerful tool to study 
the details of known reactions in given environments. 
However, interstellar/interplanetary complex organic molecule synthesis is 
complicated multi-step reactions with not well-known starting species
in poorly understood environments. 
To explore it, broad enough parameter surveys are required. 
Quantum chemistry calculation's cost would be too high 
for such surveys.

The synthesis of organic molecules (for example, sugars) in space
is a complex process in a broad range of thermal and non-thermal conditions,
while high-resolution analyses of organic molecules in carbonaceous chondrites,
UV chemistry experiments, and quantum chemistry calculations are developing.
Theoretical discussions on organic molecule synthesis in molecular clouds also have become 
active according to increasing detection of organic molecules by radio observations such as ALMA.
This situation may be similar to exoplanet studies in the 1990s and the 2000s
when the diverse mass-orbit distributions of exoplanets emerged.
At that time, we pioneered the planet population synthesis model
\citep{IL04a}, which has been substantially improved by
our and others papers \citep[e.g.,][]{IL05, Mordasini09a,IL10,IL13, Alibert13, BenzPP6}.
Multi-layered complex processes in planet formation were individually modeled
to be combined to a global sequential model, 
in order to explain the existing observational data, predict the future data,
constrain the theoretical models, and clarify missing pieces in the planet formation theory.
Although the simplified prescriptions or parameterization in the model may introduce some ambiguity, the substantially lowered calculation costs enabled broad parameter surveys.
Applying the concept of the planet population synthesis model 
for organic molecule synthesis in space,
we develop a global chemical reaction simulation scheme
that links the theory with the findings in carbonaceous chondrites and the experimental results.
Restricting our simulation to the exploration of a global picture of synthesis pathways of sugars or related molecules
in a specific environment, intermittent UV irradiation on the surface of icy particles in a protoplanetary disk,
we introduce simplifications/assumptions with 
a Monte Carlo method to select reaction sequences 
(without quantum chemistry calculations)
to accelerate the calculation substantially enough to follow global reaction
sequences of sugar synthesis and survey a broad range of parameters.
The details are explained below. 

Our new Monte Carlo chemical-reaction scheme with the graph-theoretic matrix framework 
is a forward-type method with a low computational cost.
Although the availability of our scheme has limitations in reaction conditions,
our scheme is available for the reaction sequences on icy dust particles 
intermittently irradiated by UV in warm environments that we are concerned with.
The limitations and assumptions of our model are as follows:
\begin{enumerate}
    \item We consider icy particles in a protoplanetary disk. 
    Young T Tauri stars generally emit intense FUV (6--13.6 eV) and EUV (13.6 eV or more) radiation  
    \citep[][and references therein]{Armitage2013}.
    The particles are stirred by the disk turbulence. 
    When they are occasionally pumped over the scale height,
    they are exposed to the UV flux and molecules on the icy surface 
    are photo-dissociated by UV \citep{ciesla_2012}.
    Because we consider the high-energy UV, we neglect UV absorption dependence on molecules  
    (Sect.~\ref{subsec:UV}). 
    \item We consider ``warm" regions with $T \sim$ 50--100 K 
    in the disk.  
    While only H can move on the surface of the icy particles 
    in the interstellar molecular clouds at the cold environments with $T \sim$ 10 K \citep[e.g.,][]{jin_2020}, 
    we assume that in the warm environments, molecules and
    radicals created by relatively strong UV irradiation 
    move on the surface by diffusion to react with one another
    without desorption from the surface (Sect.~\ref{weighting}). 
\end{enumerate}
Figure~\ref{fig:disk} schematically illustrates
the situation that we consider in this paper.
An icy particle at the outermost region in the protoplanetary disk
is initially small enough to be coupled with disk gas turbulence.
It is occasionally stirred by the turbulence to the upper layer to be exposed to UV radiation
from the host star.
After the particle grows to a pebble size, the particle motion is confined 
to the region near the mid-plane where it is shielded against UV radiation
\citep[also see][]{Bergner2021}.

The organic molecule synthesis on dust particles in interstellar clouds
occurs in cold ($T\sim 10$~K) and lower UV flux environments.
Diffusion of light elements on the dust surface 
would be a dominant process in the cold environments.
Because the synthesis timescale would be much longer in the lower UV flux,
interactions between the gas phase and the dust surface would play an important role.
We do not address the simulation for the interstellar clouds in this paper,
which is left for future study, because the 
main purpose of this paper is to demonstrate that our new approach has good potential
to study global features of synthesis of complex organic molecules in astronomical environments
with inexpensive computational costs.

\section{Method} \label{sec:method}

To construct our new simulation scheme,
we utilize the graph-theoretic matrix model for molecules and their chemical reactions
that was originally proposed in the 1970s by \citet{du}.
Adding a Monte Carlo scheme to the classical matrix model,
we construct a new forward-type scheme for chemical reactions for organic synthesis, 
with particular interests in the synthesis of ribose and related sugars in space. 
The classical matrix model is summarized in Sect.~\ref{subsec:matrix}.
Our new Monte Carlo scheme is explained in Sect.~\ref{weighting},
and the representation of UV irradiation is discussed in Sect.~\ref{subsec:UV}.

\subsection{A Graph-theoretic Matrix Model to Represent Molecules and Chemical Reactions}
\label{subsec:matrix}

\begin{figure*}[ht]
\begin{center}
\includegraphics[width=0.8\linewidth]{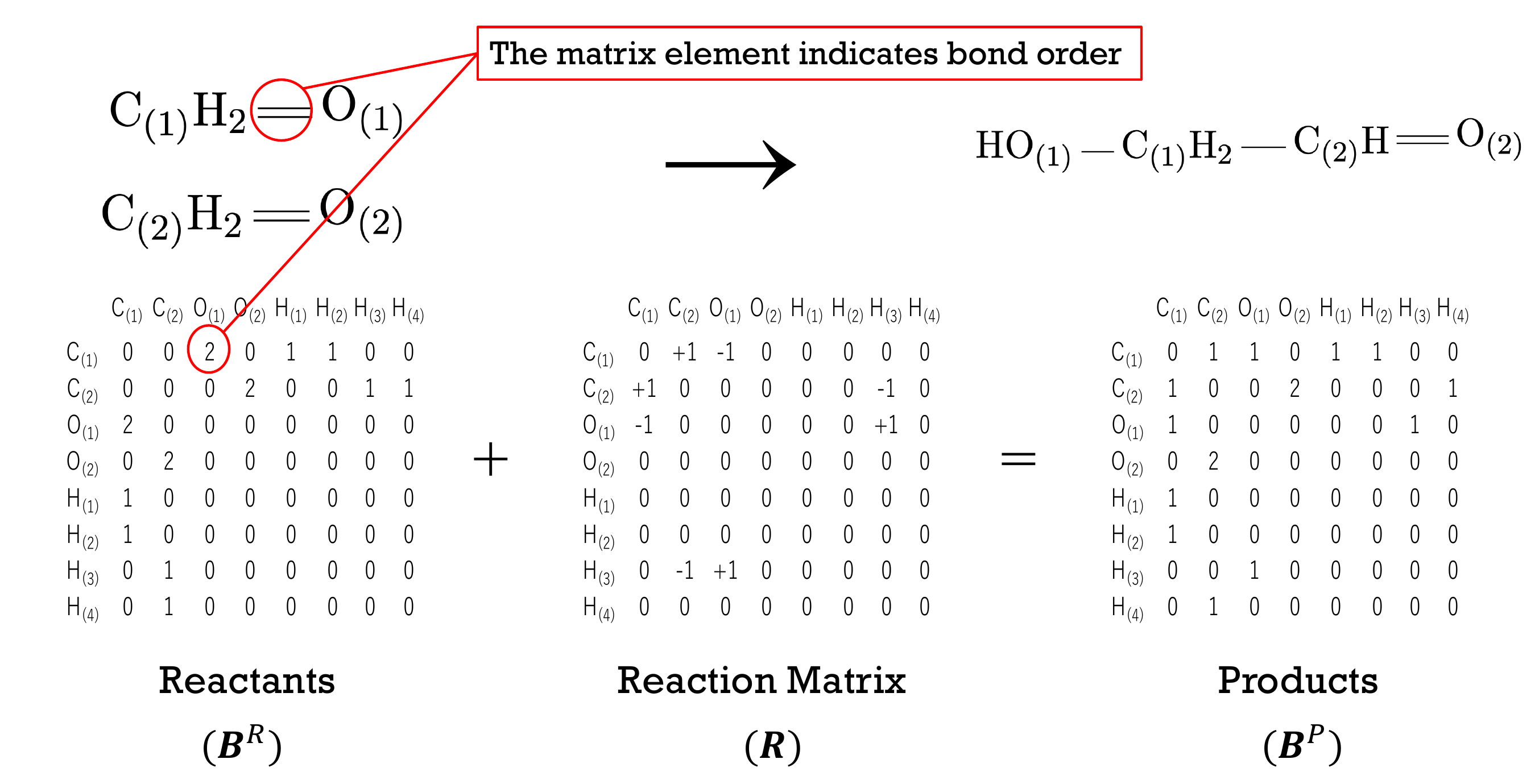}
\caption{An example of a reaction with the minimum bond change
in the matrix form (from two formaldehydes to one glycolaldehyde).}
\label{du_image}
\end{center}
\end{figure*}

A molecule with $n$ atoms can be represented by a $n \times n$ matrix $\bm{B}$; 
a non-diagonal element $B_{ij} = B_{ji}$ ($i \neq j$) indicates the bond order between the atom $A_i$ and atom $A_j$, 
and diagonal elements $B_{ij}$ ($i = j$) are always zero for covalent bonds \citep[][also see Fig.~\ref{du_image}]{du}. 
Chemical reactions can be regarded as changes in bond arrangements. 
If $\bm{B}^R$ is a bond matrix for the reactant and $\bm{B}^P$ is 
that for the product, the chemical reaction is represented by a
reaction matrix $\bm{R} = \bm{B}^P - \bm{B}^R$, which is 
an (addition) operator acting on $\bm{B}^R$.
The matrix element $R_{ij}$ indicates the change of the bond order between atom $A_i$ and $A_j$. 
Figure \ref{du_image} shows one example of our calculation, the formation of one glycolaldehyde from two formaldehydes.

In the past studies, a backward approach from a targeted product was often adopted,
and accordingly, $\bm{R}$ represented one-pod reactions with many-bond changes.
However, because we adopt a forward approach without pre-selecting targets, 
we restrict $\bm{R}$ to the minimum change with
cleavage of one existing bond and creation of one new bond;
in the matrix representation, four lines (or arrays) have only one element of $R_{ij}= -1$ 
and that of $R_{ij}= +1$ (all the other $R_{ij}$ are zero) as shown in Fig.~\ref{du_image}.
We randomly select one reaction matrix from all the possible candidates 
with weighted probabilities (Sect.~\ref{weighting})
and apply $\bm{R}$ to $\bm{B}^R$.
We regard the products $\bm{B}^P$ as the reactants $\bm{B}^R$ of the next step of the reaction sequence.
Starting from given initial molecules,  
we repeat these transformations until the reaction sequence reaches the maximum number of steps that we specify.

We modify the original matrix representation by
separating hydrogen atoms from the matrix to a vector $\bm{H}$, because
bonds with hydrogen are
uniquely determined from the non-hydrogen matrix 
with the assumption that all the bonds are covalent.
The separation of hydrogen-related bonds significantly 
reduces the matrix size and improves the calculation speed.

While the matrix representation can also treat ionic and coordination bonds, 
we only consider covalent bonds using non-diagonal elements, because
most of the organic molecules are connected with covalent bonds
and the effects of ionic bonds (electronegativity) on covalent bonds
are reflected by empirical values of bond strengths between different atoms
(Table~\ref{bond_energy} in Sect.~\ref{weighting}).
Carbon monoxide (CO) often plays an important role in chemical reactions.
We tested the inclusion of CO, using the diagonal elements, 
to find that the simulated abundance of sugars hardly changes.
Here we present the results by simpler and faster calculations without CO.
Hydrogen bonds or other non-covalent bonding
would play an important role in organic molecule synthesis in interstellar molecular clouds.
As shown in Sect.~\ref{subsec:Material_dep},
UV photo-dissociation is very efficient in the environments we consider,
the predicted abundance of sugars does not significantly depend
on starting molecule forms, but is regulated by initial atomic ratios. 
Therefore, the simulations in the most simple case in which only
covalent bonds are considered would be enough for the main purpose 
of this paper, demonstration of the availability and potential of our new model.
The inclusion of hydrogen or other non-covalent bonds is
left for future study.

\subsection{Weighting of Reaction Probability}
\label{weighting}

As summarized at the end of Sect.~\ref{sec:intro},
we adopt the following assumptions to highlight the advantage and availability of
the computationally low-cost Monte Carlo calculations: 
\begin{enumerate}
\item 
All the species interact with one another
without a restriction due to spatial distance on the surface of the icy particles.
\item Desorption from the surface is neglected (a closed system).
\end{enumerate}
Based on the above assumptions,
we survey all the possible reactions with the minimum bond change 
(Sect.~\ref{subsec:matrix}) at each step
and select one reaction from the candidate reactions 
with the following Arrhenius-type weighting:
\begin{align}
    W = \exp\left(-\frac{E_{\rm a}}{R T}\right),
    \label{eq:weight}
\end{align}
where $R$ is gas constant, $T$ is the temperature of the environment,
and $E_{\rm a}$ is an activation energy.
The energy translation is
\begin{align}
    100 \; {\rm kJ\,mol^{-1}} \simeq 1.2 \times 10^4 \, {\rm K} \simeq 1.0 \; {\rm eV}.
    \label{eq:conversion}
\end{align}
For $E_{\rm a}\sim 100 \, \rm kJ\,mol^{-1}$, the temperature dependence of $W$ is weak at $T\ga 10^4$ K.
In other words, the temperature dependence is pronounced for $T\la 10^3$ K.

The activation energy $E_{\rm a}$ can be calculated by
\textit{ab initio} density functional theory (DFT).
However, we perform calculations
of $10^5-10^6$ runs with different random numbers and each run consists of a 100 step reaction sequence with surveys of all the possible reactions at each step.
We test 26 sets of different initial species for these runs (Table~\ref{materials2}). Because DFT calculation has a high computational cost, if we use DFT calculations for the activation energy of all reactions, it significantly reduces the advantage of Monte Carlo simulation that is a global parameter survey with relatively low calculation cost.

To keep the advantage, in this work, we adopt a simpler weighting using 
the Evans-Polanyi's empirical law given by
\begin{align}
    E_{\rm a} = \alpha \Delta H + \beta \simeq \alpha \Delta D + \beta,
    \label{eq:Evans-Polanyi}
\end{align}
where $\Delta H$ is the enthalpy change during the reaction and
$\alpha$ and $\beta$ are almost constant for similar-type reactions.
The DFT calculations \citep{michaelides_2003, Wang_2011_evans, sutton_2012} suggest that
$\alpha \sim 0.6 - 1.0$ and $\beta \sim 1 - 2$~eV ($\sim 100 - 200 \, \rm kJ\,mol^{-1}$)
for a wide variety of reactions.
Because $T$ is constant during a reaction in our setting, 
$\Delta H \simeq \Delta D = D_{\rm P} - D_{\rm R}$
where $D_{\rm R}$ and $D_{\rm P}$ are the dissociation energies 
of reactants and products.
The dissociation energy is approximated by a sum of the bond energies of all pairs of atoms in the molecules.
We use the data of bond energies between
a pair of atoms given in \citet{sanderson}, which is listed in Table~\ref{bond_energy}.
For reaction selection in the Monte Carlo simulation,
the weighting factor is normalized 
by $\sum_i W_i$, the total sum of all the possible reactions ($i=1,2,..,N$), to be a probability.
Accordingly, $W \exp(\beta/RT)$ gives the same probability distribution of
the reactions as $W$ does, if $\beta$ is a constant.
Adopting $\alpha = 1$, the weighting is reduced to 
\begin{equation}
W' = W \exp \left(\frac{\beta}{R T} \right)
= \exp \left(- \frac{\Delta D}{R T} \right).
\label{reac_prob}
\end{equation}
In this paper, we use $W'$ in our Monte Carlo simulation.

We note that for the reaction to actually occur, $E_{\rm a} \la 30 \, RT$, which is equivalent to
$E_{\rm a} \la 80 \,(T/300\, \rm K) \, kJ/mol$ \citep[e.g.,][]{Rzepa2021}, is required.
As we will show in Sects.~\ref{subsec:UV} and \ref{subsubsec:sugar_synthesis},
$E_{\rm a} \ll RT$ in the UV irradiation phase in our simulations, 
and in the following non-UV phase in warm environments (``post UV phase"),
$E_{\rm a} \la 0$~kJ/mol assuming Eq.~(\ref{eq:Evans-Polanyi}),
until peaked synthesis of sugars in the post UV phase.
Thus the effect of the activation energy is negligible until the peaked sugar synthesis.
While predicted reactions after the peaked sugar synthesis 
would be affected by the activation energy, it would contribute
to longer preservation of sugars
(for details, see Sect.~\ref{subsec:E_a}).

In our prescription to calculate $\Delta D$,
where the dissociation energy is approximated by a sum of the bond energies of all pairs of atoms, 
the three-dimensional structure of molecules is not reflected.
We will discuss this issue in terms of the cyclic and chain structure of sugars
(Sect.~\ref{subsubsec:sugars_stability}).

In our simulations, we adopt UV irradiation with 10~eV, following \citet{meinert}. 
The UV energy of 10~eV is represented 
by an equivalent temperature $T = 10^5 \, \mathrm{K}$ for Eq.~(\ref{reac_prob}) (see Sect.~\ref{subsec:UV}),
while thermal room temperature ($T= 300 \,\rm K$) is assigned to 
the post UV phase to mimic the
\citet{meinert}'s experiments (see the detailed discussion in Sect.~\ref{subsec:setting}).

\begin{table}[h]
 \caption{Bond dissociation energy taken from \citet{sanderson}.
 The energy translation is
 $100 \; {\rm kJ\,mol^{-1}} \simeq 1.2 \times 10^4 \, {\rm K} \simeq 1.0 \; {\rm eV}$.
 }
 \centering
  \begin{tabular}{cc|cc}
   \hline
   & dissociation energy & & dissociation energy \\
   \hline \hline
   \ce{H - H} & 435.5 ${\rm kJ\,mol^{-1}}$ & \ce{C = C} & 610.3 ${\rm kJ\,mol^{-1}}$ \\
   \ce{C - C} & 347.0 ${\rm kJ\,mol^{-1}}$ & \ce{O = O} & 497.4 ${\rm kJ\,mol^{-1}}$ \\
   \ce{O - O} & 146.3 ${\rm kJ\,mol^{-1}}$ & \ce{C = O} & 744.0 ${\rm kJ\,mol^{-1}}$ \\
   \ce{C - H} & 413.8 ${\rm kJ\,mol^{-1}}$ & & \\
   \ce{O - H} & 464.0 ${\rm kJ\,mol^{-1}}$ & \ce{C # C} & 836.0 ${\rm kJ\,mol^{-1}}$ \\
   \ce{C - O} & 357.4 ${\rm kJ\,mol^{-1}}$ & & \\
   \hline
  \end{tabular}
  \label{bond_energy}
\end{table}

Our prescription for Monte Carlo is summarised as follows:
\begin{enumerate}
\item At each step, we list up all the possible $\bm{R}_i$ ($i=1,2,..., N$) from $\bm{B}^R$.
\item Using Table~\ref{bond_energy}, $\Delta D_i$ is calculated for all of $\bm{R}_i$.
\item The probability $W'_i$ is calculated by
Eq.~(\ref{reac_prob}) with the calculated $\Delta D_i$ and the given $T$ of the environment.
\item Generating a random number with the weighting of $W'_i$, 
a particular reaction is selected and the corresponding $\bm{R}_i$ is applied for $\bm{B}^R$ to obtain $\bm{B}^P$.
\item The product $\bm{B}^P$ is assigned to $\bm{B}^P$ for the next step and we go back to 1.
\end{enumerate}

\subsection{UV Irradiation}
\label{subsec:UV}

In our simulations, we adopt UV irradiation with 10~eV, following \citet{meinert}. 
The UV radiation energy of 10~eV corresponds to an equivalent temperature 
$\sim 1.2 \times 10^5 \,\rm K \sim 10^3\,\rm kJ\,mol^{-1}$ (Eq.~(\ref{eq:conversion})).
Because UV irradiation with 10~eV would result in the bond cleavage to create radicals 
rather than transition to a higher bond energy level,
the wavelength dependence of UV absorption for individual molecules may not be sensitive
\citep[e.g.,][]{Limao2003}
and the representation of UV irradiation by $T=10^5 \, \mathrm{K}$ 
would be a good approximation.
While the radicals may stay on the icy surface, we assume that they can diffuse and 
interact with any other species on the surface
\citep[for liquid-like behavior of ice radiated by UV in this temperature range, see][]{tachibana_2017}.

Because $E_{\rm a}$ and $\Delta D$ for a minimum bond change are typically $< 500 \, \rm kJ\,mol^{-1}$,
which is equivalent to $< 6\times 10^4$~K,
for the minimum bond change reactions that we consider (see also Table~\ref{bond_energy}),
the weighting $W$ given by Eq.~(\ref{eq:weight}) 
and $W'$ given by Eq.~(\ref{reac_prob}) are $\sim 1$ for all of possible reactions
under the UV irradiation.
In other words, the energy of 10~eV UV irradiation, corresponding to
$\sim 10^5$~K, is sufficiently higher than the energy barriers of each reaction, and thus the activation energy is negligible in the UV irradiation phase.


In our fiducial simulation runs, we calculate $N_{\rm U} = 70$ steps in the UV phase
followed by $N_{\rm p} = 30$ steps at 300~K without UV irradiation
(Sect.~\ref{subsec:setting}).
While steps in the 300 K phase represent time evolution, 
the interpretation of the UV phase needs to be careful.
In our simulations, radicals are not explicitly treated.
We formally assume that atoms are in covalent bonds,
but their bonds are actually cleaved by UV.
What we simulate is the equilibrium distribution of molecules
that temporarily exist as very loosely bounded or 
that is expected to establish when UV irradiation flux decays.
In this sense, UV phase calculation is equivalent to setting-up of 
initial conditions for the post-UV low-temperature evolution.

\subsection{Parameter Settings}
\label{subsec:setting}

\begin{table}[ht]
 \caption{Patterns of a set of initial conditions: species, H/O, and H/C ratios of starting materials. P1 is the fiducial set.}
  \label{materials2}
 \centering
  \begin{tabular}{lccc}
   \hline
   & starting materials & H/O ratio & H/C ratio\\
   \hline \hline
   P1 & 7 $\mathrm{CH_2O}$, 20 $\mathrm{H_2O}$ & 2.00 & 7.71 \\
   \hline
   P2 & 4 $\mathrm{CH_2O}$, 3 $\mathrm{CO_2}$, 4 $\mathrm{H_2}$ & 1.60 & 2.29 \\
   P3 & 5 $\mathrm{CH_2O}$, 2 $\mathrm{CO_2}$, 9 $\mathrm{H_2O}$ & 1.56 & 4.00\\
   P4 & 2 $\mathrm{CH_3OH}$, 5 $\mathrm{CO_2}$, 17 $\mathrm{H_2O}$ & 1.45 & 6.00\\
   P5 & 2 $\mathrm{CH_2O}$, 5 $\mathrm{CO_2}$, 25 $\mathrm{H_2O}$ & 1.46 & 7.71\\
   P6 & 7 $\mathrm{CO_2}$, 32 $\mathrm{H_2O}$, 3 $\mathrm{H_2}$ & 1.52 & 10.00 \\
   P7 & 7 $\mathrm{CH_2O}$ & 2.00 & 2.00 \\
   P8 & 13 $\mathrm{CH_2O}$, 14 $\mathrm{H_2O}$ & 2.00 & 4.15 \\
   P9 & 10 $\mathrm{CH_2O}$, 17 $\mathrm{H_2O}$ & 2.00 & 5.40 \\
   P10 & 7 $\mathrm{CH_2O}$, 17 $\mathrm{H_2O}$ & 2.00 & 6.86 \\
   P11 & 7 $\mathrm{CH_2O}$, 24 $\mathrm{H_2O}$ & 2.00 & 8.86 \\
   P12 & 7 $\mathrm{CO_2}$, 21 $\mathrm{H_2O}$, 14 $\mathrm{H_2}$ & 2.00 & 10.00 \\
   P13 & 7 $\mathrm{CH_2O}$, 17 $\mathrm{H_2O}$, 3 $\mathrm{H_2}$ & 2.25 & 7.71 \\
   P14 & 7 $\mathrm{CH_2O}$, 4 $\mathrm{H_2O}$, 3 $\mathrm{H_2}$ & 2.55 & 4.00 \\
   P15 & 7 $\mathrm{CH_2O}$, 10 $\mathrm{H_2O}$, 4 $\mathrm{H_2}$ & 2.47 & 6.00\\
   P16 & 7 $\mathrm{CH_2O}$, 14 $\mathrm{H_2O}$, 6 $\mathrm{H_2}$ & 2.57 & 7.71 \\
   P17 & 7 $\mathrm{CH_2O}$, 18 $\mathrm{H_2O}$, 6 $\mathrm{H_2}$ & 2.48 & 8.86 \\
   P18 & 7 $\mathrm{CH_3OH}$, 20 $\mathrm{H_2O}$ & 2.52 & 9.71 \\
   P19 & 7 $\mathrm{CH_3OH}$, 24 $\mathrm{H_2O}$ & 2.45 & 10.86 \\
   P20 & 7 $\mathrm{CH_2O}$, 4 $\mathrm{H_2}$ & 3.14 & 3.14\\
   P21 & 7 $\mathrm{CH_3OH}$, 7 $\mathrm{H_2O}$ & 3.00 & 6.00\\
   P22 & 7 $\mathrm{CH_3OH}$, 11 $\mathrm{H_2O}$, 3 $\mathrm{H_2}$ & 3.11 & 8.00 \\
   P23 & 7 $\mathrm{CH_2O}$, 16 $\mathrm{H_2O}$, 5 $\mathrm{H_2}$ & 3.04 & 10.00\\
   P24 & 7 $\mathrm{CH_2O}$, 1 $\mathrm{H_2O}$, 6 $\mathrm{H_2}$ & 3.50 & 4.00 \\
   P25 & 7 $\mathrm{CH_2O}$, 5 $\mathrm{H_2O}$, 9 $\mathrm{H_2}$ & 3.50 & 6.00 \\
   P26 & 7 $\mathrm{CH_2O}$, 9 $\mathrm{H_2O}$, 12 $\mathrm{H_2}$ & 3.50 & 8.00 \\
   \hline
  \end{tabular}
\end{table}

While our primary purpose is to 
investigate how complex organic molecules are generally synthesized on the surface of icy dust particles 
by intermittent UV irradiation in the protoplanetary disks,
we define fiducial parameters in our simulations as
those corresponding to \citet{meinert}'s experiments.
This is because quantitative comparisons with the experiments would be 
important to construct and constrain the theoretical model.  
\citet{meinert} suggested formation of ribose by formose-type reactions.
However, they used methanol as starting materials, rather than 
formaldehyde because of its difficulty of treatment.
In numerical simulations, such difficulty does not exist and we use 7 formaldehyde molecules and 20 water molecules (7 CH$_2$O + 20 H$_2$O) as fiducial starting materials
to compare the suggested formose-type reactions with the sugar synthesis paths
obtained by our numerical simulation.
In Sect.~\ref{subsec:Material_dep},
we will show that a specific choice of starting molecule structures
does not affect the peak abundance of ribose in the post UV phase.

In organic photochemistry experiments, as in \citet{meinert}, 
the samples are warmed up to room temperature after UV irradiation for the spectroscopic analysis.
Because the warmed-up phase may induce important reactions for sugar synthesis,
we consider two phases in our simulation, 
the UV phase ($T=T_{\rm U}$) and the post UV phase ($T=T_{\rm p}$).
Because \citet{meinert} used 10~eV UV and set the sample at room temperature for the analysis, 
we define fiducial temperatures as $T_{\rm U} =10^5 \, \rm K$ and $T_{\rm p} =300\, \rm K$.

In Sect.~\ref{sec:intro}, we mentioned that our simulation setting
would be justified for $T_{\rm p} \sim$~50--100~K.
We show in Sect.~\ref{subsec:Temp_dep} that
the predicted sugar abundance is almost the same for $T_{\rm p} \la 10^3$~K
in our simulations.
Our simulation setting that assumes preservation of all of the atoms
breaks down for the room temperature or more, because
volatile molecules such as H$_2$O sublimate.
However, the complex organic molecules would remain to be analyzed.
Accordingly, we set $T_{\rm p} =300\, \rm K$ in most of our simulations.

For each set of initial molecules, 
we repeat $n_{\rm run}$ = $10^5$--$10^6$ runs of $N_{\rm U} = 70$ step reactions in the UV phase 
and $N_{\rm p} = 30$ step reactions in the post UV phase.
In each run, a random number sequence is different.
We will show that the results are not affected by the choices of $T_{\rm U}$ and $T_{\rm p}$ 
as long as $T_{\rm U} \ga 10^4\, \rm K$ and $T_{\rm p} \la 10^3\, \rm K$.
As shown in Sect.~\ref{sec:result}, 
the sugar synthesis does not depend on $N_{\rm U}$ and $N_{\rm p}$ as long as
$N_{\rm U} \ga 40$ and $N_{\rm p} \ga 20$.
For safety, we adopt $N_{\rm U} = 70$ and $N_{\rm p} = 30$.

We tested 26 sets with different starting materials that are summarized in Table~\ref{materials2}.
We first show the results of simulations with the fiducial initial materials and parameters in detail (set P1).
After that, we investigate the dependence of sugar synthesis on temperature and starting materials. 
We compare the predicted relative abundance of sugars, sugar alcohols, and deoxysugars
with the experimental results.

\section{Results} \label{sec:result}

\begin{figure}[ht]
\begin{center}
\includegraphics[width=8cm]{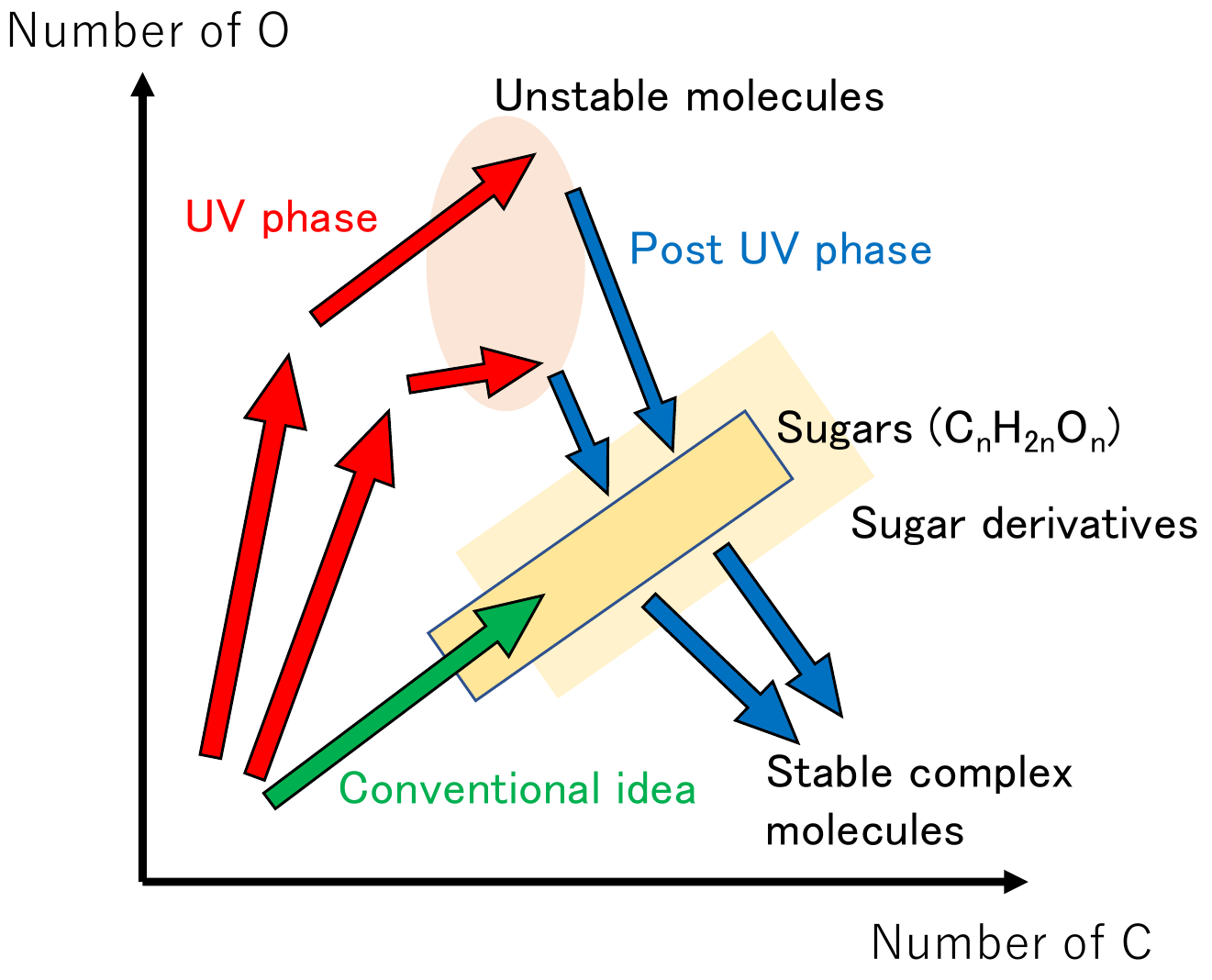}
\caption{Sugar synthesis pathways shown by our simulation 
(the red and blue arrows) and by the conventional formose-type reactions
(the green arrows).
The horizontal and vertical axes are the numbers of carbon and oxygen atoms contained in molecules, respectively. }
\label{mechanism}
\end{center}
\end{figure}

The synthesis pathway of sugars found in our simulations 
is the formation of large O-rich molecules randomized by UV irradiation
followed by irreversible break-ups of the O-O bonds to be
replaced by C-O/O-H and replacements of -OH by -H
in the post UV phase.
It is a very different pathway from the conventional
step-by-step bottom-up pathway such as formose-type reactions
\footnote{For the synthesis of complex organic molecules in interstellar clouds,
a scenario of formation of large molecules by radical-radical interactions
followed by sputtering and photofission was also proposed
\citep[e.g.,][]{Sorrell2001}, although
UV and temperature environments are different from those in the present paper.}. 
Figure~\ref{mechanism} is a schematic illustration of the two pathways.
Below, we show the details of every part
of the pathway that we find, based on the detailed numerical results.
The schematic illustration helps understand the whole picture of the pathway.

\subsection{Simulation for Fiducial Parameters}
\label{result1}

\begin{figure}[ht]
\begin{center}
\includegraphics[width=8cm]{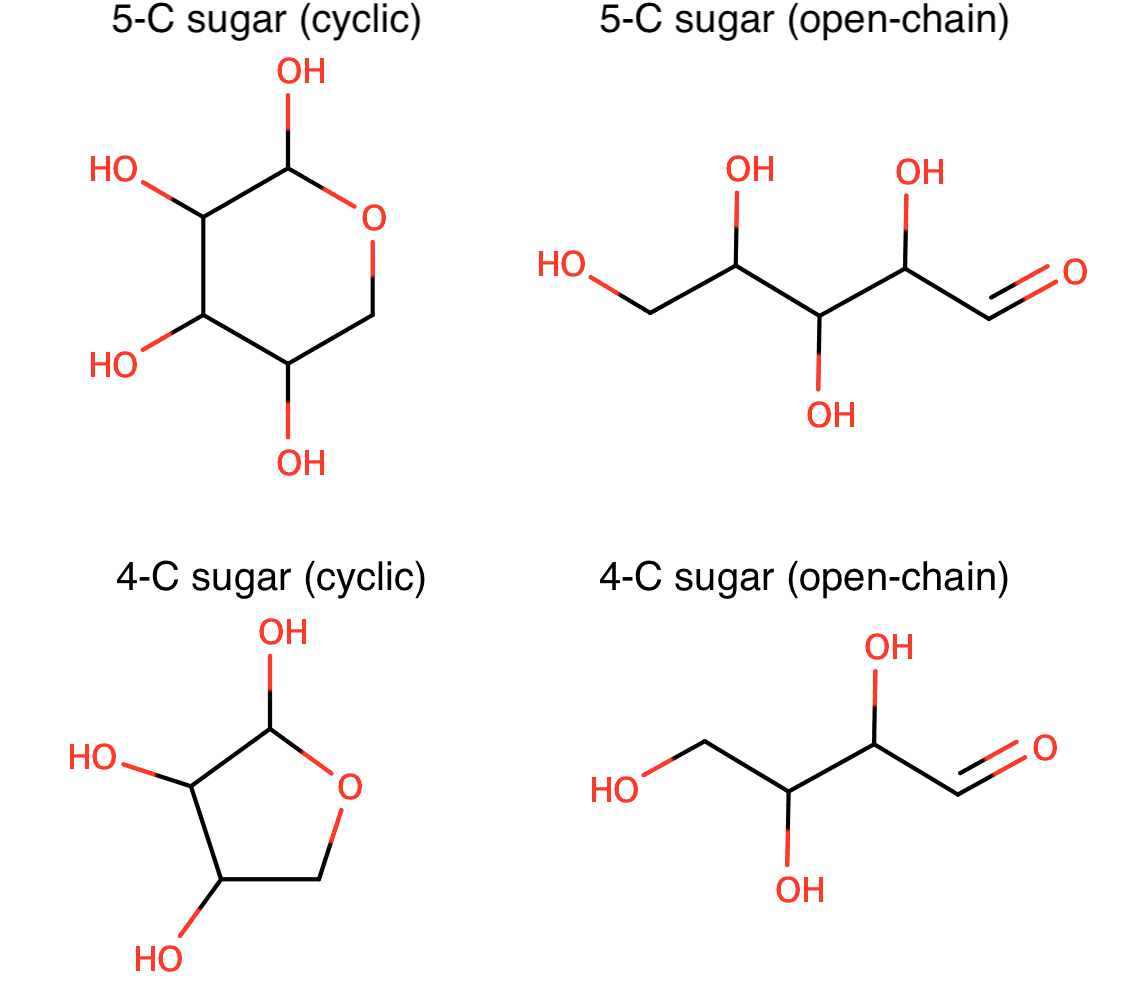}
\caption{Chemical structures of {\it k}-C sugars ($k=4,5$) that we count to calculate the abundance.
We count both the (pyranose) cyclic and open-chain forms.}
\label{sugar_structure}
\end{center}
\end{figure}

We show the abundance evolution of 4-C and 5-C sugars
as a function of a reaction step.
In this paper, the ``abundance" is defined as the frequency for targeted sugar molecules to exist at each step.
If targeted molecule {\it k}-C sugars exist 
at the $n_{\rm step}$-th step in $n_{\mbox{$k$-C}}$ runs out of the total runs $n_{\rm run}$,
the abundance at $n_{\rm step}$ is given by 
\begin{align}
    P_k = n_{\mbox{$k$-C}}/n_{\rm run}.
    \label{eq:abundance}
\end{align}
Here ``sugars" that we count are the molecules with the cyclic or open-chain structure shown in Fig. \ref{sugar_structure}.
We do not count {\it k}-C sugars in the furanose form ($k$-membered ring), because most of $k$-C sugars existing in the cyclic form have pyranose structure ($(k+1)$-membered ring) and the definition should be simple to avoid ambiguity. 

\begin{figure}[ht]
\begin{center}
\includegraphics[width=8cm]{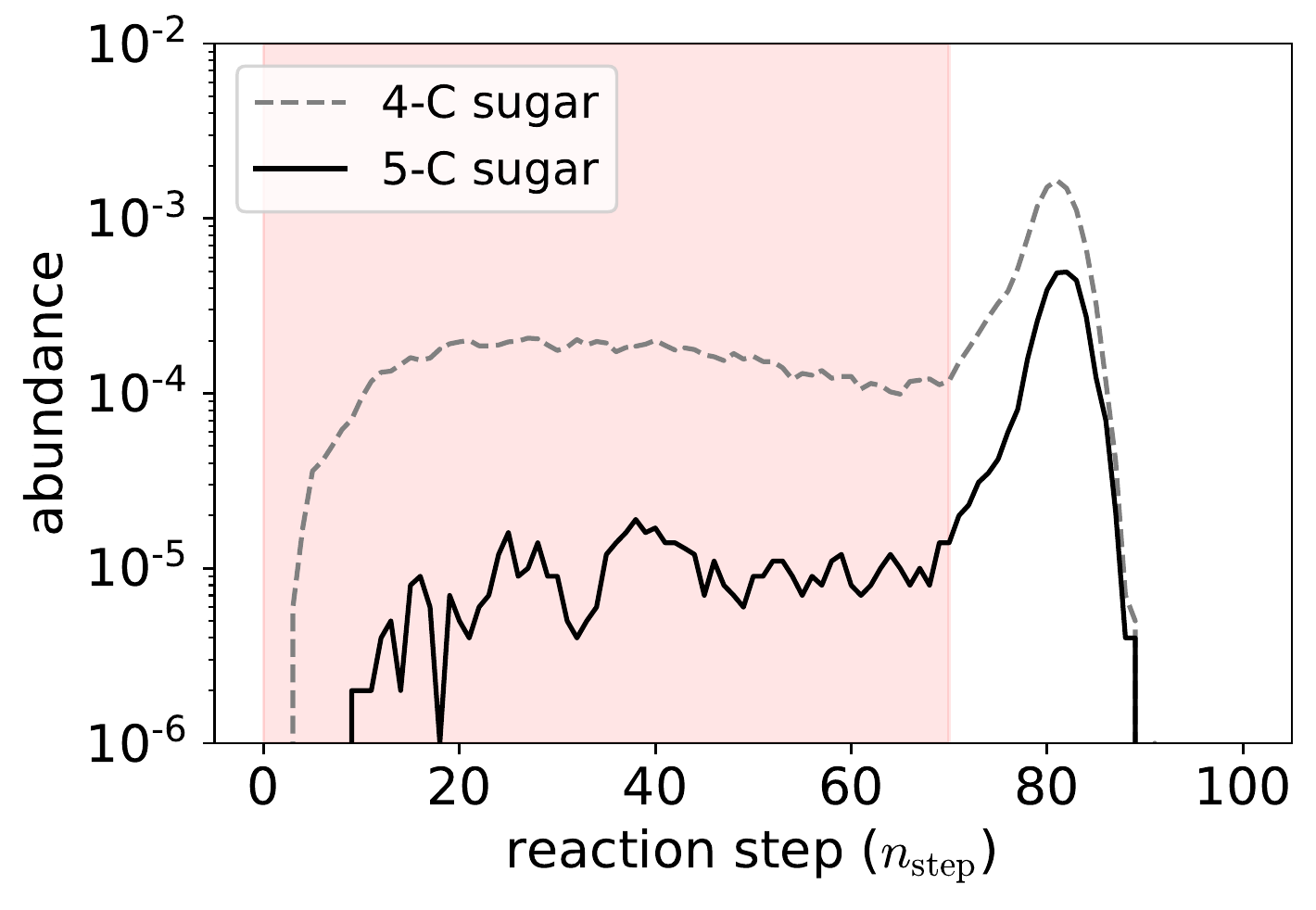}
\caption{The abundances of 4-C and 5-C sugars as a function of a reaction step for the fiducial set (P1),
starting from 7 $\mathrm{CH_2O}$ and 20 $\mathrm{H_2O}$, and temperature of $T_{\rm U} = 10^5$ K and $T_{\rm p} = 300$ K. 
The abundance is defined by Eq.~(\ref{eq:abundance}). 
The grey dashed line and black solid line are for 4-C and 5-C sugars, respectively. 
The red area corresponds to the UV phase.}
\label{result_fp}
\end{center}
\end{figure}

Figure \ref{result_fp} shows the results
for fiducial parameters: starting from 7 $\mathrm{CH_2O}$ and 20 $\mathrm{H_2O}$, 
$T_{\rm U} = 10^5$ K (10~eV of UV irradiation) and $T_{\rm p} = 300$ K.
The total runs are $n_{\rm run}=10^6$.
Synthesis of 4-C and 5-C sugars begin after a few and $\sim 10$ steps,
respectively, and their abundance keeps an almost constant value 
from $n_{\rm step} \sim 30$--40 through the end of UV phase ($n_{\rm step} = 70$).
In the post UV phase at 300 K ($71 \le n_{\rm step} \le 100$),
the abundance of 4-C and 5-C sugar rapidly increases until $n_{\rm step} \sim 80$ followed by quick decay.
As we will discuss, the final rapid decay would not be realistic, while the evolution until
the peaked sugar synthesis may be actually realized.

\subsubsection{Large Molecules Formed in the UV Phase}
\label{subsubsec:UV}

\begin{figure*}[ht]
\begin{center}
\includegraphics[width=0.8\linewidth]{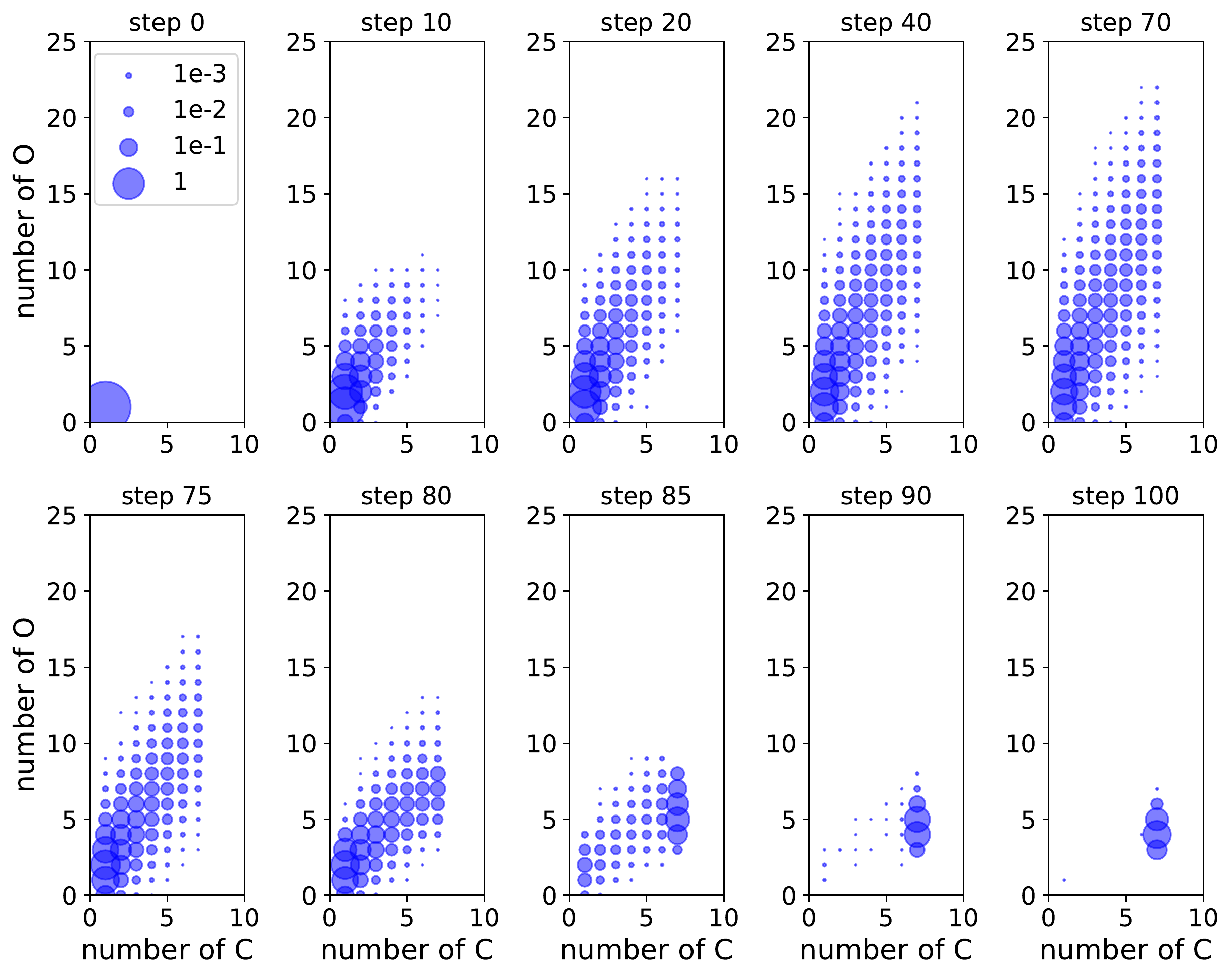}
\caption{Evolution of the molecule distribution for the fiducial parameters. 
The horizontal and vertical axes indicate the number of C and O in the individual molecules at each step. 
All of the molecules with different numbers of C and O at each step 
are plotted with the abundance (out of the total runs $n_{\rm run}=10^5$)
represented by the circle sizes.
The upper and lower panels show the evolution in the UV phase and the post UV phase,
respectively.
}
\label{evolution}
\end{center}
\end{figure*}

As we discussed in Sect.~\ref{subsec:UV},
the simulated evolution in the UV phase does not necessarily mean actual time evolution,
but it rather shows the relaxation to 
the equilibrium distribution of molecules
that is regarded as the initial conditions for the post UV low-temperature evolution.
In the fiducial case, thanks to the initially existing formaldehyde, 
4-C and 5-C sugars are quickly formed.
However, because the equivalent energy with $T_{\rm U} = 10^5$ K is higher than 
the bond energy changes associated with the minimum bond change reactions
($\la 350 \, {\rm kJ\,mol^{-1}}$; Table~\ref{bond_energy}),
the weights of Monte Carlo calculations,  
$\mathrm{exp} (-{\Delta D}/828 \, \rm kJ\,mol^{-1})$ for $T_{\rm U} = 10^5$~K,
are $\simeq$ 1 for all of the possible
minimum bond change reactions from the molecules at each step.
Accordingly, UV irradiation randomly rearranges chemical bonds by repeated cleavage and combination of all possible reactions, independent of their bond strength.
As a result, UV generates a variety of molecule sizes.

Figure~\ref{evolution} shows the distribution of molecules 
at each step on the plane of C and O atom numbers of the molecules. 
The circle sizes represent the abundance.
Because seven C atoms initially exist, the distribution is truncated at $\rm C=7$.  
While small molecules also exist, 
the distribution spreads and large molecules are synthesized as $n_{\rm step}$ proceeds, 
which have some chain or branched structures consisting of the sequences of bonds of carbon and oxygen atoms.
For $n_{\rm step} \ga 40$ in the UV phase, the distribution reaches an equilibrium.
Sugar molecules are also formed in this phase, 
but they are decomposed immediately due to random rearrangements of bonds.
For $n_{\rm step} \ga 40$ in this UV phase,
these repeated cleavage and combination balance and the sugar abundance are kept almost constant.

\subsubsection{Sugar Synthesis in the Post UV Phase}
\label{subsubsec:sugar_synthesis}

\begin{figure}[t]
\begin{center}
\includegraphics[width=8cm]{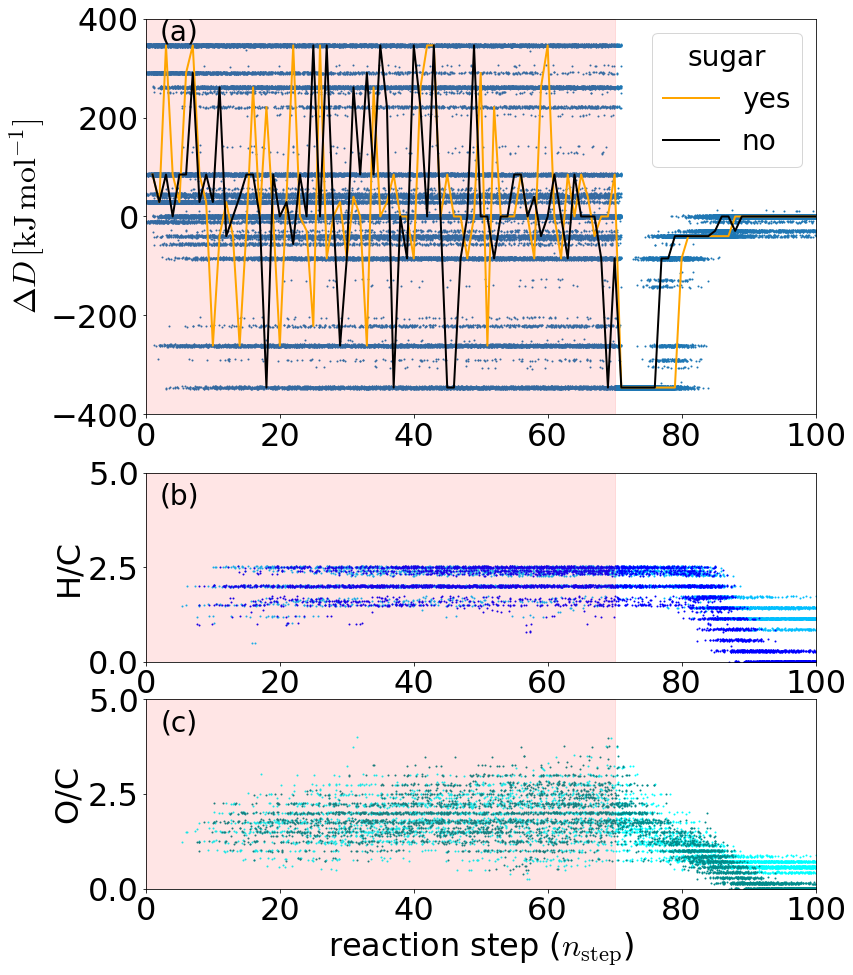}
\caption{The dissociation energy and associated composition changes of large molecules
along reaction sequences in the UV phase ($n_{\rm step} = 1$--70)
and the post UV phase ($n_{\rm step} = 71$--100).
The top panel is (a) the distribution of $\Delta D$ selected by the Monte Carlo simulations
at each step in $10^3$ runs in set P1. 
The horizontal axis is a reaction step, and the vertical axis is $\Delta D$ in the unit of $\rm kJ\,mol^{-1}$. 
The yellow and black lines are typical reaction sequences where 5-C sugar is formed
and that without sugar formation, respectively.
The middle and bottom panels show the distributions of (b) H/C and (c) O/C ratios of the molecules with four or more carbon atoms
at each step in the $10^3$ runs.
The dark color dots indicate the results in the case allowing the formation of three or four-membered ring structures, and light color dots are the results with prohibiting their formation.
For visual convenience, the dots with the same $\Delta D$, H/C, and O/C at $n_{\rm step}$ are shifted slightly so that they do not overlap and clustered dots represent frequent events.}
\label{dE}
\end{center}
\end{figure}

After UV irradiation is turned off and the $T$ is switched to 300 K ($n_{\rm step} \ge 71$), 
the Monte Carlo calculation is drastically changed.
The weight becomes $\mathrm{exp} (-{\Delta D}/2.48\, \rm kJ\,mol^{-1})$
and reactions are sensitively contrasted by the weights.
Accordingly, the abundance of 4-C and 5-C rapidly increases, as explained below.

The dissociation energy of O-O bonds is
smaller than that of C-O bonds by $\sim$210 $\rm kJ\,mol^{-1}$ 
and that of O-H bonds by $\sim$320 $\rm kJ\,mol^{-1}$ (Table~\ref{bond_energy}).
For the reaction including this bond change,
$\Delta D$ is usually $\sim -(260$--350) $\rm kJ\,mol^{-1}$
and the reaction probability is higher than the other reactions
by a huge factor. 
Therefore, the reaction to break O-O bonds are predominantly selected
by the Monte Carlo calculation, and the existing O-O bonds 
are replaced by the stronger bonds one after another until all of the O-O bonds are broken.
Molecules are rapidly converted to more stable structures. 

The top panel of Fig.~\ref{dE} shows $\Delta D$ at each reaction step.
The blue dots are the distributions of $10^3$ runs in set P1.
A typical path of $\Delta D$ for a reaction sequence of 5-C sugar synthesis and that without sugar formation are also shown.
In a few steps after the termination of the UV irradiation ($n_{\rm step} = 70$),
only the reaction with $\Delta D = - 346\, \rm kJ\,mol^{-1}$ is selected.
The reaction is the break-up of O-O bonds by attacks of $\rm H_2$:
$\rm H_2$ + O-O $\rightarrow$ 2 OH (Table~\ref{bond_energy}).
For $n_{\rm step} \sim 75$--85, 
other reactions to break up O-O bonds, C-H + O-O $\rightarrow$ C-O + O-H
with $\Delta D = -261\, \rm kJ\,mol^{-1}$ and H-C-O-O $\rightarrow$ C=O + O-H with $-291\, \rm kJ\,mol^{-1}$, 
are also selected, in addition to the above reaction.
The reaction of replacing C-OH by a stronger bond C-H with
the formation of H$_2$O ($\Delta D = -85\, \rm kJ\,mol^{-1}$) also starts at $n_{\rm step} \sim 75$ and
continues through the end of the runs. 
As the break-ups of O-O bonds proceed, the molecule sizes decrease 
as shown in the panels
$n_{\rm step} = 70, 75, 80$ and 85 in Fig.~\ref{evolution}.

\begin{figure*}[ht]
\begin{center}
\includegraphics[width=0.8\linewidth]{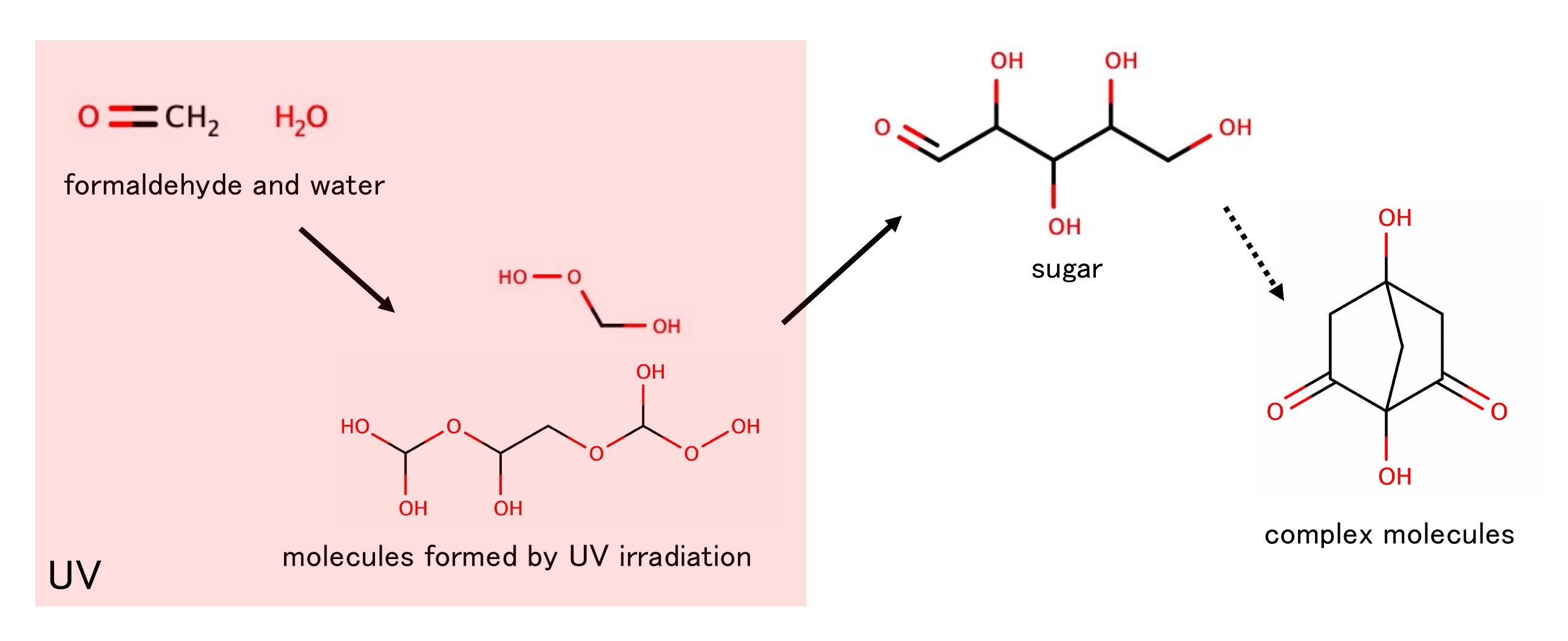}
\caption{One example of reaction paths of ribose synthesis. 
The red-shaded region represents the evolution during the UV phase
and the non-shaded region shows the evolution in the post UV phase.}
\label{path}
\end{center}
\end{figure*}

In this phase, the abundance of 4-C and 5-C sugars rapidly increases 
until $n_{\rm step} \sim 80$, as shown in Fig.~\ref{result_fp}. 
In the post UV phase, large unstable molecules with $\rm O/C > 1$ formed in the UV phase
are decomposed by the break-ups of O-O bonds,
irreversibly decreasing the O/C ratio
and probabilistically producing sugars.

Figure~\ref{dE}~c shows the evolution of O/C of molecules
composed of four carbon atoms or more.
During the UV phase, O/C values are distributed
broadly at $\sim$1.0--3.0.
Because H/C is distributed at $\sim$1.5--2.5, as shown in 
Fig.~\ref{dE}~b, a small fraction of the molecules
is in the state of sugars, $\rm O/C = 1$ and $\rm H/C = 2$.
Soon after the UV irradiation is turned off, 
the O/C distribution is converged with the decrease of its mean value
through the break-ups of O-O bonds and
the replacements of -OH by -H, 
while the H/C distribution is kept constant. 
As a result, molecules 
arrive at sugars with $\rm O/C = 1$ and $\rm H/C = 2$
with small but non-negligible probability ($\sim 5 \times 10^{-4}$). 
Figure \ref{path} shows one example of sugar synthesis pathways.

The Evans-Polanyi's empirical law with DFT calculations 
(Eq.~\ref{eq:Evans-Polanyi}) suggests that
the activation energy can be neglected for the reactions with
$E_{\rm a} \la 80 \,(T/300\, \rm K) \, kJ/mol$ (Sect.~\ref{weighting}).
For $\alpha = 1$, this condition corresponds to
$\Delta D \la [80 \,(T/300\, \rm K) \, kJ/mol - \beta)]$,
which is $\sim -20$ to $-120 \, \rm kJ/mol$ at $T \simeq 300 \, \rm K$.
Because the reactions including the O-O bond break-ups have
$\Delta D \sim -350$ to $-260\, \rm kJ\,mol^{-1}$,
it is likely that sugar synthesis by this process is not inhibited by its activation energy.
On the other hand, the replacements of -OH by -H with $\Delta D \sim -85 \, \rm kJ\,mol^{-1}$ 
may be marginal to occur.

In Sect.~\ref{subsec:Temp_dep}, we confirm that no sugar is synthesized 
in the case where $T=300$ K is set in all steps.
The formation of large O-rich molecules at $T \ga 10^4$ K (corresponding to the energy $\ga$ 1 eV) is a necessary condition for the sugar synthesis in our setting.
Thus, UV radiation plays an important role in the enhanced synthesis of sugars in the post UV phase.

\subsubsection{Preservation of Sugars}
\label{subsubsec:sugars_stability}

In our simulations, 4-C and 5-C sugar abundance is peaked at $\sim$80 steps.
After the peak, the replacements of -OH by -H 
continue and molecules become further O-poor ($\rm O/C < 1$), as shown in 
Fig.~\ref{dE}~c.
Because $\rm O/C = 1$ for sugars,
the sugar abundance rapidly decreases.
After the peak, H/C also starts decreasing through reactions
of replacing C-H bonds by C-C bonds, because H is taken by O to form 
H$_2$O with high bond strength.
In the case where three or four-membered ring structures are allowed
(the dark color dots),
we find more O-poor ($\rm O/C \ll 1$) molecules are formed.
We will come back to this final evolution that could be related to
formation of IOM in Sect.~\ref{subsec:IOM}.  

Because the reaction of replacing -OH by -H has $\Delta D = -85 \, \rm kJ\,mol^{-1}$
in our prescription (Table~\ref{bond_energy}), 
the activation energy may not be low enough to be neglected
and it is affected by the three-dimensional structure of molecules. 
Accordingly, we need a more careful discussion on the decomposition of sugars found in our simulations.

Sugars in a solution have two types of forms, the open-chain and the cyclic forms.
It is empirically well known that the cyclic form is more stable than the open-chain form, 
and most sugars are in the cyclic form in a solution.
Because sugars change a form in a solution,
sugars can occasionally take the open-chain form and be decomposed,
although the probability of the decomposition is small.
As a result, the decomposition timescale is relatively long at room temperature. 
\citet{dass_cyclic} suggests that the cyclic form has lower energy by $\sim$10--30 $\rm kJ\,mol^{-1}$
than the open-chain form and \citet{azofra_2012} suggests $\sim$50 $\rm kJ\,mol^{-1}$.
Because the two forms have similar dissociation energies in our prescription,
the above energy difference should reflect the three-dimensional structure, which we do not take into account.
The accurate value of this conversion energy is not clear enough
for us to explicitly introduce to our simulation.
The reaction of replacing -OH by -H has $\Delta D = -85 \, \rm kJ\,mol^{-1}$
and is marginal for the decomposition due to the activation energy.
The additional energy decrease of $\sim$10--50 $\rm kJ\,mol^{-1}$ for the cyclic form
may result in the preservation of sugars against the -OH/-H replacement reactions.

In this paper, we adopt a stable-limit assumption that 
sugars formed at room temperature is quickly transformed into the more stable
cyclic form to inhibit further decomposition by the -OH/-H replacements.
Thereby, the peak values of sugar abundance are used to compare with the experimental results (Sects.~\ref{subsec:Meinert} 
and \ref{subsec:deoxysugar}).

The large molecules prepared in the UV phase are
decomposed into smaller molecules with fewer oxygen atoms by the
break-up of O-O bonds and the -OH/-H replacements through many different paths.
On the way of the reduction of oxygen atoms in molecules,
a small fraction of the molecules happen to pass sugars and may be stabilized by
the transformation to the cyclic form.
In Fig.~\ref{dE}~a, we also plot the paths in which 5-C sugars are formed in the post UV phase as well as those without sugar synthesis.
It shows that the paths are similar to each other, independent of whether sugars are synthesized or not,
which suggests that the sugar synthesis is just by chance following the probability distribution.
A more detailed preservation mechanism of synthesized sugars is left for future work.

\subsection{Dependence on Temperature}
\label{subsec:Temp_dep}

\begin{figure*}[ht]
  \begin{minipage}{0.5\linewidth}
    \centering
    \includegraphics[keepaspectratio, 
    width=8cm]{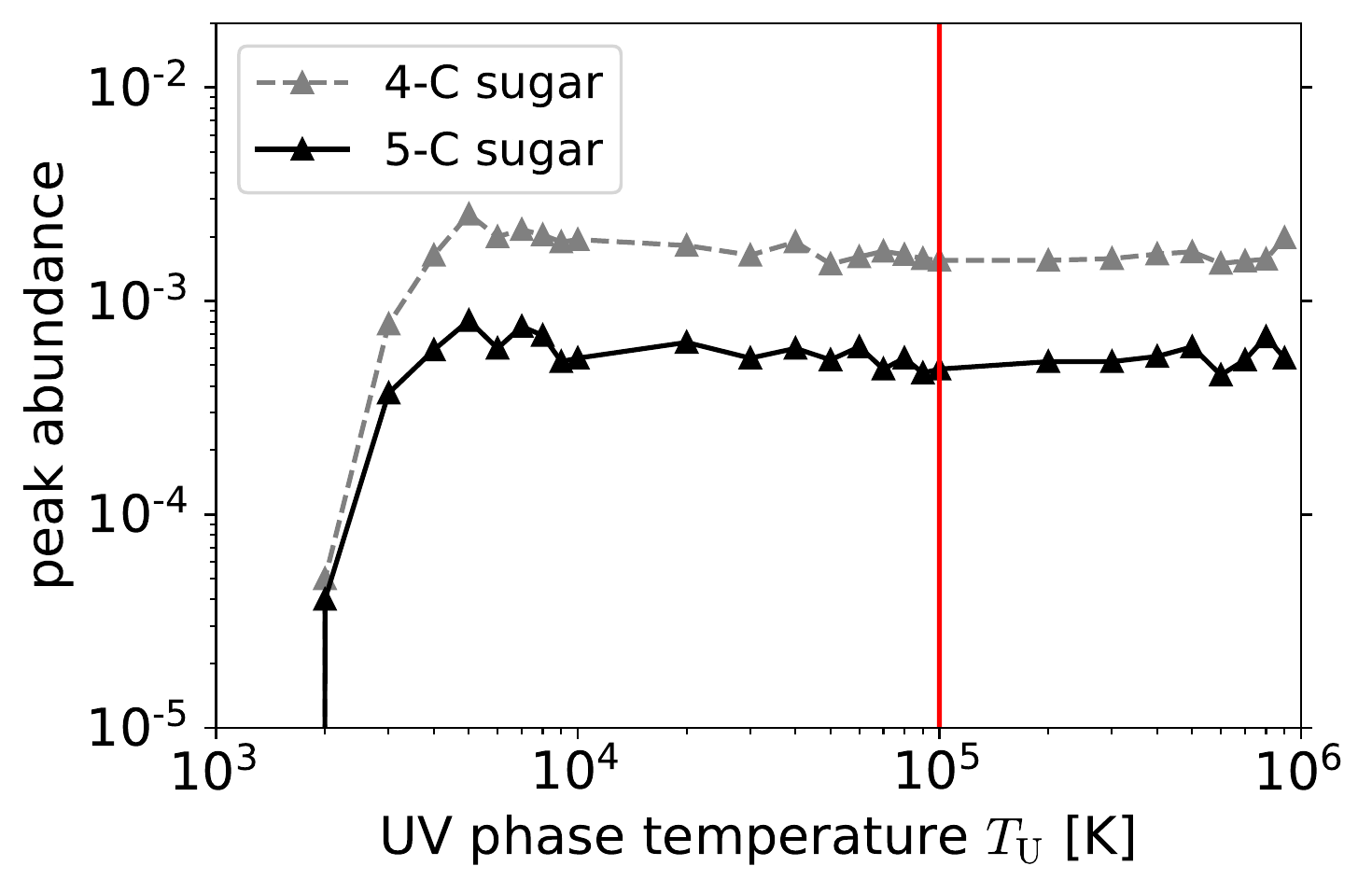}
    \subcaption{}
    \label{tu}
  \end{minipage}
  \begin{minipage}{0.5\linewidth}
    \centering
    \includegraphics[keepaspectratio, 
    width=8cm]{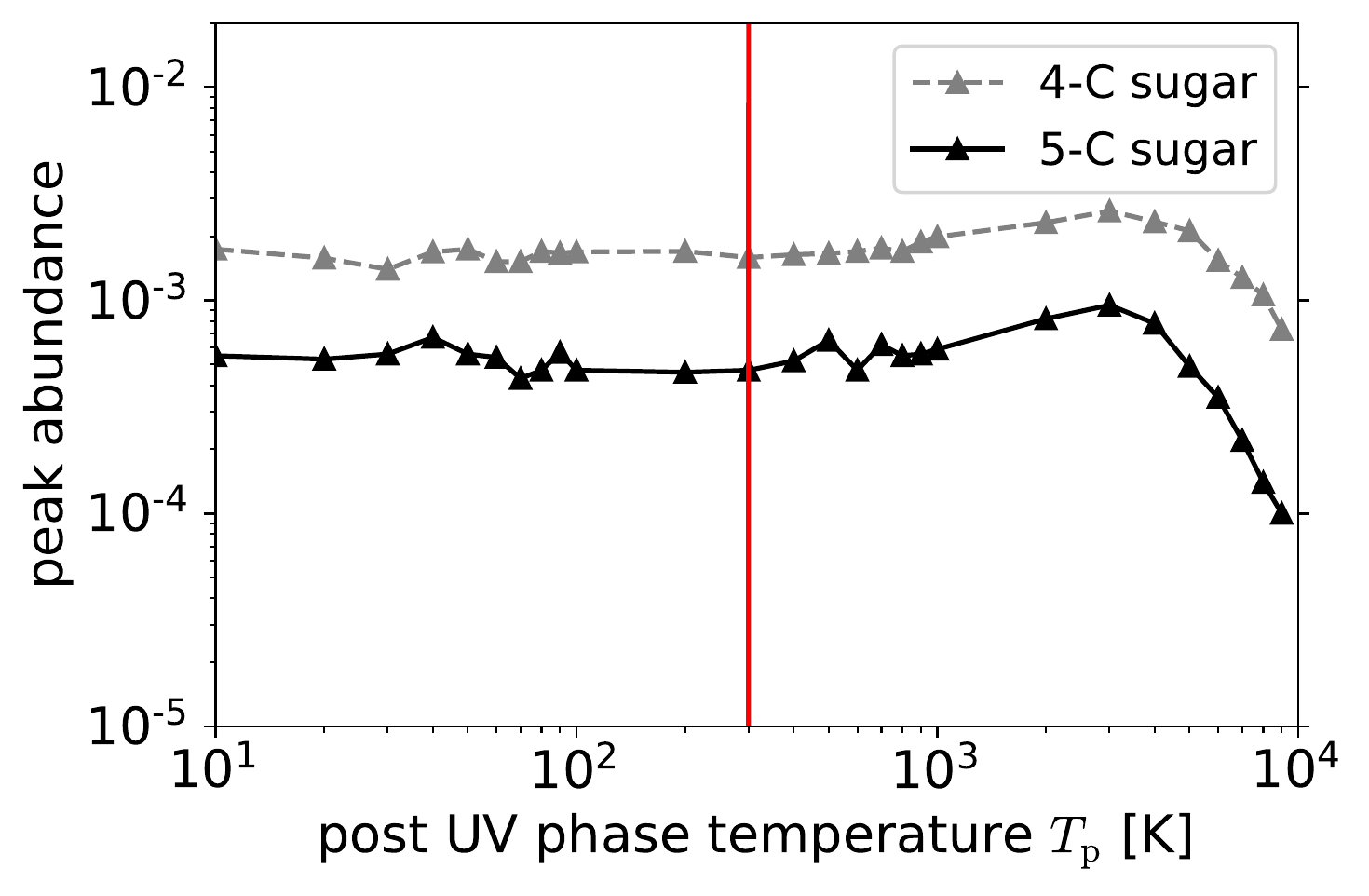}
    \subcaption{}
    \label{tn}
  \end{minipage}
 \caption{The peak sugar abundances in the post UV phase for different temperatures when $n_{\rm run}=10^5$.  We use fiducial starting materials. The left panel (a) shows the dependence on temperature of the UV phase when $T_{\rm p} = 300$ K, and the right panel (b) shows the dependence on temperature of the post UV phase when $T_{\rm U} = 10^5$ K. The grey dashed line and black solid line are for 4-C and 5-C sugars, respectively. Red lines indicate the positions of the fiducial temperatures.}
  \label{temp_dependence}
\end{figure*}

In the fiducial runs, we set $T_{\rm U}=10^5 \, \rm K$ (10~eV of UV irradiation) and $T_{\rm p}=300 \, \rm K$. 
Here, we show the peak sugar abundance with different $T_{\rm U}$ and $T_{\rm p}$.
Figure~\ref{temp_dependence}~a shows the peak values of 4-C and 5-C sugar abundances in the post UV phase as a function of $T_{\rm U}$
for the fixed $T_{\rm p} = 300 \, \rm K$.
The peak abundances of 4-C and 5-C sugars are 2 $ \times 10^{-3}$ and 5--6 $ \times 10^{-4}$, respectively, independent of $T_{\rm U}$ for $T_{\rm U} \ga 10^4$~K. 
Figure~\ref{temp_dependence}~b shows the dependence on $T_{\rm p}$ for the fixed $T_{\rm U} = 10^5 \, \rm K$.
Both abundances are the same for $T_{\rm p} \la 10^3$~K.

The typical bond energy difference $|\Delta D|$ is $\sim$~10--350~$\rm kJ\,mol^{-1}$ (Fig.~\ref{dE}~a),
which corresponds to the temperature of about $10^3$--$3 \times 10^4$~K.
When $T_{\rm U} \ga 10^4$~K, most of the reactions occur without a large difference in the probability
and a similar distribution of molecules is prepared for the irreversible reactions
in the post UV phase.
On the other hand, when $T_{\rm p} \la 10^3$~K, 
which corresponds to $\la 10\, \rm kJ\,mol^{-1}$, 
the reactions with $\Delta D \sim -350$ to $-260\, \rm kJ\,mol^{-1}$
are continuously selected if the unstable O-rich molecules are remained.
Thereby, the rapid increase of sugar abundance after UV irradiation is 
rather a universal event in a broad range of the environment temperature, 
but not a tuned event that occurs only at a specific temperature.

\subsection{Dependence on Starting Materials}
\label{subsec:Material_dep}

\begin{figure}[ht]
\begin{center}
\includegraphics[width=8cm]{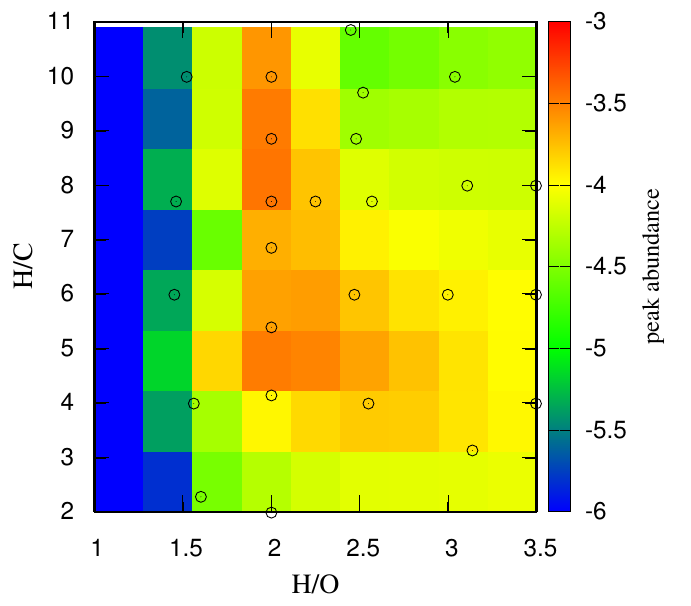}
\caption{The peak value of sugar abundance in the post UV phase for different element ratios of starting materials. The horizontal axis is the H/O ratio, and the vertical axis is the H/C ratio. The color bar indicates the peak value with a logarithmic scale; the red color means a large value.}
\label{ho_hc_ratio}
\end{center}
\end{figure}

If sugars are synthesized by a step-by-step bottom-up pathway such as formose-type reactions,
starting species would be important.
However, in the sugar synthesis pathways that we have shown here,
the equilibrium molecule distribution is established by the random bond arrangements by UV irradiation.
We find that the peak abundance of sugars is regulated by
atom ratios such as H/O and H/C, but not by specific starting species.

In order to understand how sugar abundance depends on starting materials, 
we performed runs in sets with a variety of starting materials listed in Table~\ref{materials2}.
The results are summarized in Fig.~\ref{ho_hc_ratio}.
The circles indicate H/O and H/C ratios of starting materials we tested,
and the color contours indicate the peak value of 5-C sugar abundance in the post UV phase
interpolated by the results with discrete values of H/O and H/C ratios.
We find three important features.

\begin{figure}[ht]
\begin{center}
\includegraphics[width=8cm]{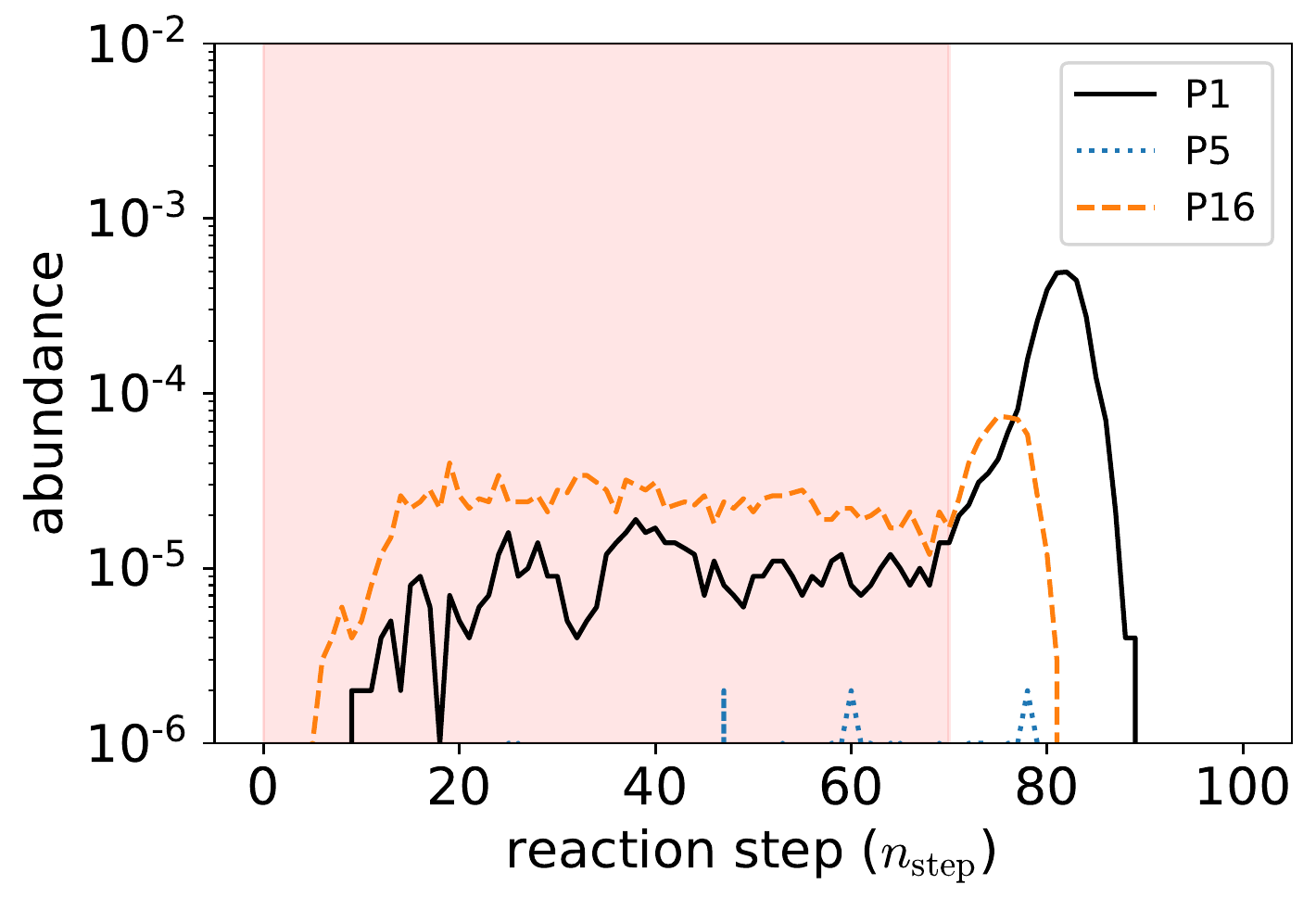}
\caption{Abundance of 5-C sugar as a function of a reaction step in the simulations for P1, P5, and P16 when $T_{\rm U} = 10^5$ K, $T_{\rm  p} = 300$ K and $n_{\rm run} = 10^6$. The black solid, blue dotted, and orange dashed lines represent the results for P1, P5, and P16, respectively.}
\label{result_ho}
\end{center}
\end{figure}
The first one is that the sugar synthesis efficiency
is mostly regulated by the H/O ratio.
The peak value is obtained with H/O $\simeq$ 2.0 for a fixed H/C.
As we showed, molecules convert to more stable structures with strong bonds in the post UV phase.
Figure~\ref{result_ho} shows the sugar abundance for three patterns (P1, P5, and P16 in Table \ref{materials2}).
These patterns have the same H/C ratio (= 7.71) and different H/O ratio 
(H/O = 2.00, 1.46, and 2.57 for P1, P5, and P16).
The peak value in the post UV phase is the largest for H/O $\simeq$ 2.
On the other hand, few sugars are synthesized in both the UV phase and the post UV phase for H/O $\simeq$ 1.5 (O-rich environment; P5).
In this case, oxygen atoms are abundant at the initial state 
and most of the molecules formed in the UV phase include more oxygen than sugars, 
so that sugar synthesis is not efficiently induced in the UV phase.
In the post UV phase, most carbon atoms combine with oxygen and they construct molecules without hydrogen, such as carbon dioxide, rather than carbon chains necessary for sugar synthesis, because C=O bonds are very strong.
When the H/O ratio is high (H-rich environment; P16), in the post UV phase, most carbon atoms construct small carbohydrate molecules and oxygen atoms tend to form water molecules because the C-H bond is stronger than the C-O bond.
It demonstrates that too oxidizing or reducing environment is not suitable for sugar synthesis.

\begin{figure}[ht]
\begin{center}
\includegraphics[width=8cm]{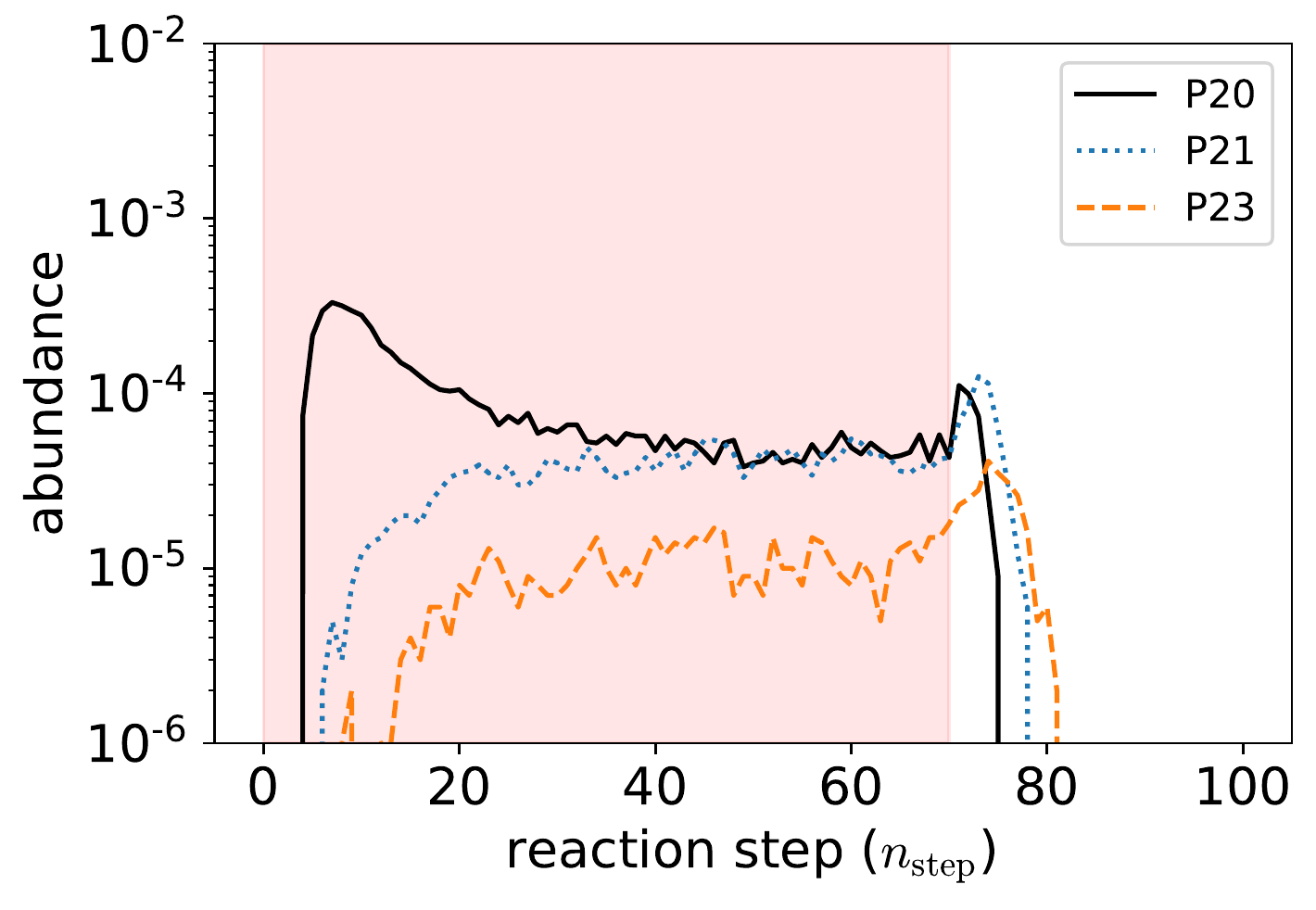}
\caption{Same as Fig.~\ref{result_ho} except for P20, P21, and P23 instead of P1, P5 and P16.}
\label{result_hc}
\end{center}
\end{figure}
Second, for H/O $> 2$, the abundance decreases as the H/C ratio increases
from $\rm H/C \simeq 5$.
Figure~\ref{result_hc} shows the results of P20, P21, and P23
that have the same H/O ratio ($\simeq$ 3) but different H/C ratios; 
H/C = 3.14, 6.00, and 10.00 for P20, P21, and P23, respectively.
The peak value of the sugar abundance in the post UV phase is smaller for high H/C ratios (P23).
We also note that the sugar abundance in the early UV phase is different, although it does not depend significantly on H/O
(Fig.~\ref{result_ho}).
For high H/C ratios, excess hydrogen atoms combine with carbon atoms and produce small hydrocarbons such as methane, 
preventing the molecules from becoming large with many carbon atoms.
Even in the post UV phase, hydrogen atoms break C-C bonds because C-H bonds are stronger than C-C bonds, inhibiting the formation of carbon chains that constitute sugars.

Finally, the abundance also decreases at $\rm H/C \la 2$ even if H/O $\simeq$ 2.
When both H/C and H/O are low, carbon atoms tend to combine with each other, leading to the generation of open or cyclic carbon chains, and most of the large molecules existing in the UV phase contain many carbon atoms.
After UV irradiation is finished, 
they keep the main structures because they already have many C-C bonds and few C-H bonds, and only oxygen atoms bonded to carbon chains are separated. 
Accordingly, they do not pass sugar molecules with $\rm O/C = 1$.

\subsection{Comparison with \citet{meinert}'s Experiment}
\label{subsec:Meinert}

\citet{meinert} mentioned that it is not clear when the ribose synthesis occurred in their experiments, 
in which they perform UV irradiation at 78 K followed by the transfer to the room temperature environment.
Our simulation in this paper suggests that the dominant synthesis occurred
at the room temperature after UV irradiation, although 
the synthesis pathway in our simulation (formation of O-rich molecules 
during UV irradiation followed by the break-ups of O-O bonds
and replacements of -OH by -H at the room temperature) is 
totally different
from formose-type building-up reactions suggested by \citet{meinert}
in most cases.

\begin{figure}[ht]
\begin{center}
\includegraphics[width=8cm]{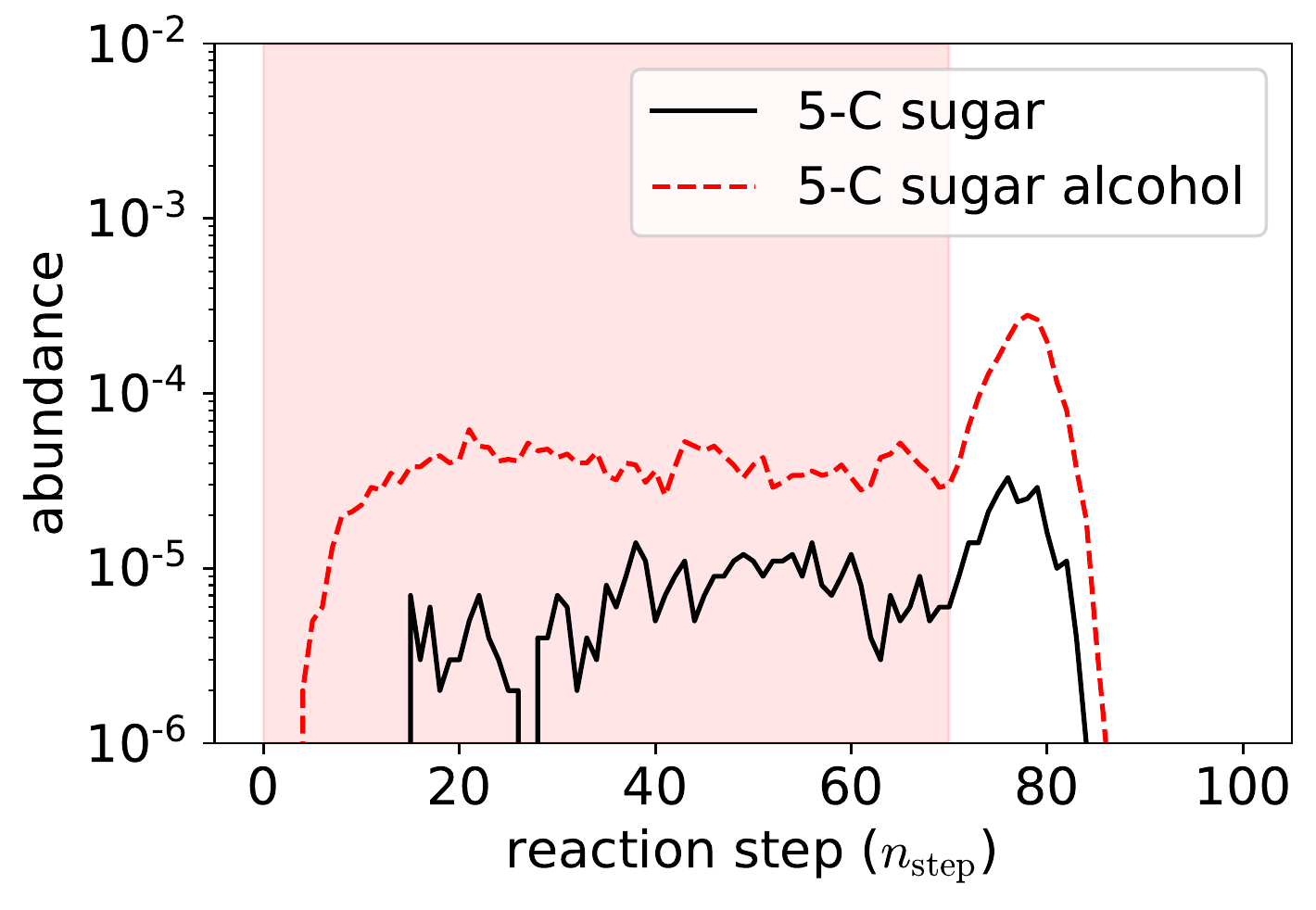}
\caption{Abundances of 5-C sugars and 5-C sugar alcohols as a function of a reaction step for set P5. 
The black solid and red dashed lines are for 5-C sugar and 5-C sugar alcohol, respectively.}
\label{alcohol}
\end{center}
\end{figure}

Because \citet{meinert}~also reported sugar alcohols (derivatives of sugars) in their products, we also examine the abundance of sugar alcohols
in the simulation with initial species of 7 CH$_3$OH (methanol) and 20 H$_2$O that is similar to their experiments.
We set $T_{\rm U}=10^5$~K and $T_{\rm p}=300$~K.
The result is shown in Fig.~\ref{alcohol}.
The abundance ratio of 5-C sugar alcohols to 5-C sugars 
is $R_{\rm al} \simeq 6.2$ at the peak in the post UV phase
($n_{\rm step} = 78$).
The ratio found in \citet{meinert}'s experiment is
$R_{\rm al} \simeq 4.1$, which is comparable to our simulation result.

\subsection{Synthesis of Deoxysugars}
\label{subsec:deoxysugar}

\begin{figure}[ht]
\begin{center}
\includegraphics[width=8cm]{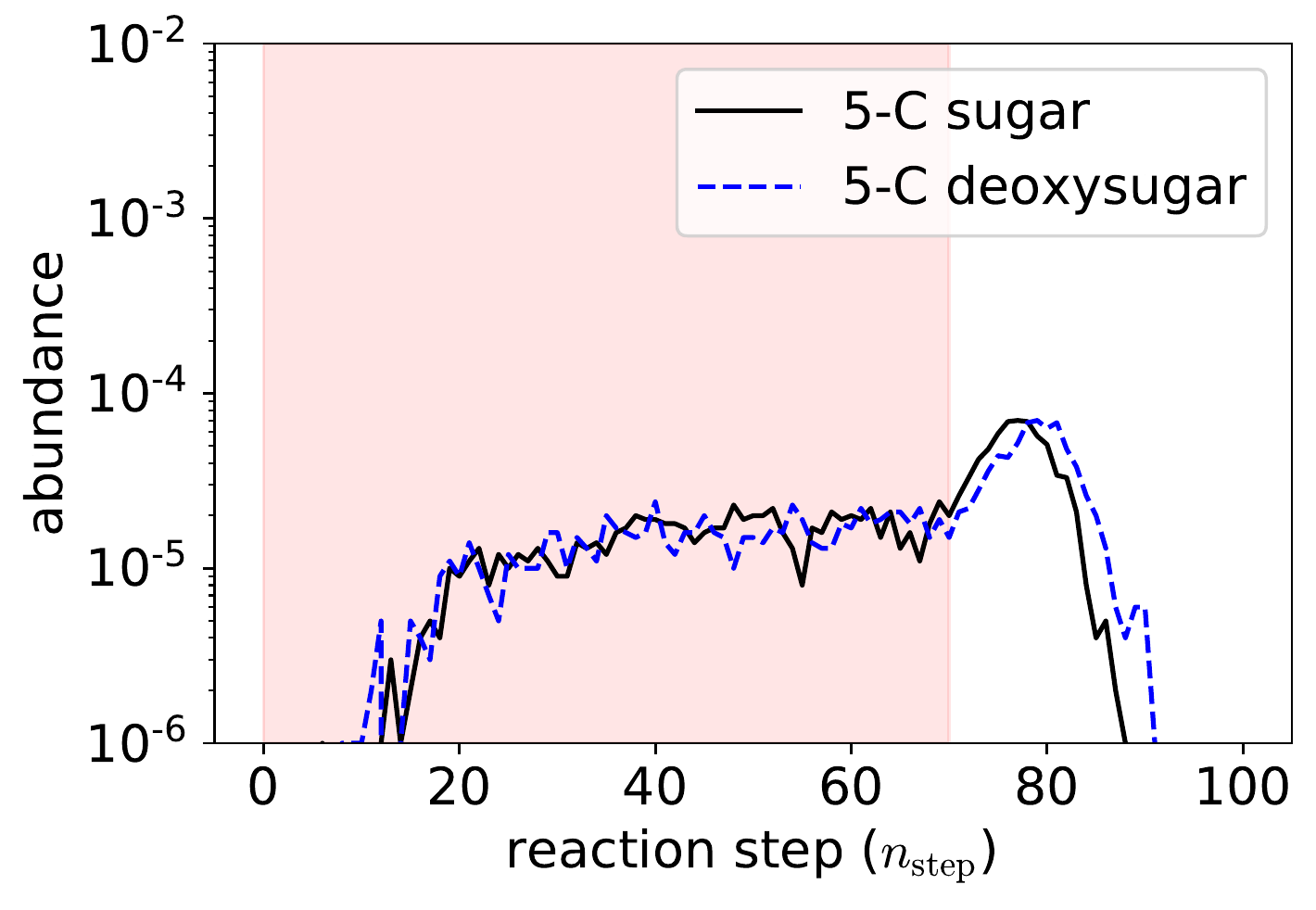}
\caption{Abundances of 5-C sugars and 5-C deoxysugars as a function of a reaction step. 
The black solid line and blue dashed line indicate 5-C sugar and 5-C deoxysugar, respectively.}
\label{deoxy}
\end{center}
\end{figure}

Another experiment on sugar synthesis by UV irradiation \citep{nuevo} reported that deoxysugars were detected from the products and 
the abundance ratio of 5-C deoxysugars to 5-C sugars is 
$R_{\rm dxy} \sim 1$.
The abundance ratio in our simulation starting from 
10 CH$_3$OH (methanol) and 20 H$_2$O,
which is similar to \citet{nuevo}'s experiment,
with $T_{\rm U}=10^5$ K and $T_{\rm p}=300$ K shows
$R_{\rm dxy} \sim 1$ in entire steps (Fig.~\ref{deoxy}).

It is known that deoxysugars are not formed via formose-type reactions \citep{butlerow1861, breslow1959}.
The detection of their derivatives, such as deoxysugar alcohols, in meteorites was reported \citep{cooper2001, nuevo}.
Possible step-by-step building-up type pathways of deoxysugars synthesis 
were suggested in previous studies \citep{oro1962, ritson2014}.
While the previously proposed synthesis pathways 
for sugars and deoxysugars are independent of each other, 
the pathway shown in our simulations can simultaneously produce sugars and deoxysugars.
As we argued in Sect.~\ref{subsubsec:sugar_synthesis},
in our simulations, the sugar synthesis occurs  
on the way of decomposition process at 300 K 
from large O-rich molecules generated during the UV phase to
C-rich molecules.
Our results show that sugars are synthesized not by a single unique path but by 
probabilistic, divergent paths. 
In other words, the synthesis paths have a ``broad" range in phase space,
compared with the conventional formose-type reactions.
Because the structure of 5-C deoxysugars is similar to 5-C sugars
and their synthesis probabilistically occurs,
it is reasonable that their abundance is always similar to each other,
almost independent of environmental conditions.
The broad range enables the simultaneous synthesis of sugars, deoxysugars, and other related molecules.
Because the cyclic form of deoxysugars would also be more stable,
the same argument of preservation may be applied for deoxysugars.

\section{Discussion} \label{sec:discussion}

\subsection{Activation Energy}
\label{subsec:E_a}

In the Monte Carlo simulation in the present paper, 
we use the weighting factor $W' = \exp(- \Delta D/RT)$ (Eq.~(\ref{reac_prob})) that does not depend on 
the activation energy $E_{\rm a}$, instead of the more accurate weighting
with $W = \exp(- E_{\rm a}/RT)$ (Eq.~(\ref{eq:weight})).
During the UV phase, $E_{\rm a} \ll RT$ with $T \sim 10^5$~K
and all the minimum bond change reactions occur with the same probability,
independent of $E_{\rm a}$ (Sect.~\ref{subsec:UV}).
In the post UV phase where UV irradiation is turned off and $T=300$~K is set, 
reactions of breaking up unstable large molecules with   
$\Delta D \sim -350$ to $-260\, \rm kJ\,mol^{-1}$ are repeated 
until sugar synthesis is peaked (Sect.~\ref{subsubsec:sugar_synthesis}).
The Evans-Polanyi's empirical law, $E_{\rm a} \simeq \alpha \Delta D + \beta$
(Eq.~(\ref{eq:Evans-Polanyi}))
where $\alpha \sim 0.6$--1 and $\beta \sim 100$--$200 \rm \, kJ\,mol^{-1}$, suggests that
$E_{\rm a}$ is negligible ($E_{\rm a}$ is $\la 0$~kJ/mol), 
which is too small to induce a energy barrier in this case.
The unstable O-rich large molecules are formed by UV irradiation
and they play an important role in enhanced synthesis of sugars/deoxysugars
in the post UV phase.
Therefore, as long as we are concerned with 
the peaked abundance of sugars/deoxysugas, 
the effect of the activation energy should be negligible
and it is justified to use $W'$ as the weighting factor.

However, as noted in Sect.~\ref{weighting}, 
for the reaction to actually occur, $E_{\rm a} \la 80 \rm \, kJ/mol$ is required at 300~K, which is equivalent to $\Delta D \la -20$ to $-120 \rm \, kJ/mol$ for $\alpha \sim 1$.
For the decomposition reactions of sugars/deoxysugars by replacement of -OH by -H,
$\Delta D \simeq -85 \rm \, kJ\,mol^{-1}$.
The reactions after the peaked synthesis of sugars/deoxysugars have $\Delta D$ with smaller absolute values (Fig.~\ref{dE}).
It is not clear the reactions after the peaked sugars/deoxysugars synthesis 
proceeds at 300~K.
This could contribute to the longer preservation of sugars/deoxysugars.
The assumptions of our simulations are consistent with the ``warm" environment of
$T \sim 50$--100~K. 
In this case, the sugars/deoxysugars preservation should be further longer.

In summary, 
\begin{enumerate}
    \item UV phase: 
    
    The effect of $E_{\rm a}$ is negligible by the high
    equivalent $T$ of UV irradiation ($E_{\rm a} \ll RT$).
    
    \item Early post UV phase until the peaked sugar/deoxysugar synthesis:

    The effect is negligible for the repeated 
    reactions with $\Delta D \sim -350$ to $-260\, \rm kJ\,mol^{-1}$ ($E_{\rm a} \la 0$). 

    \item Later post UV phase after the peaked synthesis:

    The effect
    cannot be neglected, but its inclusion would prolong the preservation of sugars/deoxysugars, which is 
    rather more consistent with our arguments.
\end{enumerate}
Therefore, the effect of $E_{\rm a}$ would not significantly change the results in this paper. 

To study the effect of the activation energy in more details, we need to 
perform quantum chemistry calculations 
or use chemical reaction data bases
\footnote{In our prescription, any reaction is divided into
the sum of the minimum bond change reactions and
the transition state is partially taken into account in some cases. 
For example, ammonia formation from N$_2$ is divided into three reactions, 
$\rm N_2 + 3H_2 \rightarrow N_2H_2 + 2H_2 
\rightarrow 2NH_2 + H_2 \rightarrow 2NH_3$.
For the whole reaction of $\rm N_2 + 3H_2 \rightarrow 2NH_3$, $\Delta H < 0$.
However, for the first and second reactions, $\Delta H > 0$,
implying that the divided states of $\rm N_2H_2 + 2H_2$ and $\rm 2NH_2 + H_2$ 
express energy barriers.}.
Because key reactions in the later post UV phase are limited, it may not be too difficult,
although we leave it for future study.

We also point out in Sect.~\ref{subsubsec:sugars_stability} that
the transformation sugars/deoxysugars from the open-chain form to the cyclic form
may contribute to the preservation of sugars/deoxysugars.
More detailed discussions on bond energy originated from the three-dimensional structure of the molecules,
which we do not take into account in the present paper, are also needed.


\subsection{Synthesis of IOM-like Molecules}
\label{subsec:IOM}

The mechanism found by our simulation simultaneously synthesizes sugars and deoxysugars.
It could be consistent with the co-existence of sugars and deoxysugar derivatives in meteorites and 
the co-synthesis of sugars and deoxysugars found by a photochemistry experiment \citep{nuevo}.

On the other hand, our simulation (Fig.~\ref{dE}~a) suggests that
most of the molecules, which do not pass the sugars or deoxysugars, 
repeatedly suffer the reaction of 
C-H + C-OH $\rightarrow$ C-C + H$_2$O with 
$\Delta D \simeq - 40\, \rm kJ\,mol^{-1}$ (Table~\ref{bond_energy}). 
It decreases H/C at $n_{\rm step} \ga 80$ (Fig.~\ref{dE}~b),
in addition to the steady decrease in O/C in the entire period of the post UV phase (Fig.~\ref{dE}~c).
The final products are large C-rich molecules
\footnote{When hydrogen excess is substantial, some carbon atoms are combined with hydrogen atoms and produce methane, which is mentioned in Section \ref{subsec:Material_dep}.}.

In Fig.~\ref{dE}~b, the light and dark color dots
represent the result with and without the prohibition of three or four-membered ring structures, respectively.
In general, three or four-membered ring structures are unstable,
because the wave functions of the atoms in the ring overlap.
In our simulations, the formation of three or four-membered ring structures is prohibited.
Following this prescription, $\rm H/C \rightarrow 1.7$ and $\rm O/C \rightarrow 0.86$
(the light color dots).
However, we find that due to the limitation of only 7 carbon atoms in the simulations with
the prohibition of three or four-membered ring structures,
complex carbon rings with several -OH groups remain.
In reality, carbon atoms are almost infinite, and
more C-rich molecules with smaller H/C and O/C are likely to be formed.
We also performed runs with the same initial molecules and 
the same random number sequences without the
prohibition of three or four-membered ring structures (the dark color dots).
Until the sugar synthesis, the distributions of H/C and O/C
are almost the same as those with the prohibition.
After the sugar synthesis, however, more C-rich final products with lower H/C and O/C
are formed.
The final molecules have multiple combinations of three and four-membered ring structures.
In reality, stable molecules with multiple combinations of five and six-membered ring structures would be formed.

These theoretically inferred final products have structures
similar to the complex Insoluble Organic Matter (IOM)
that is a major component in organic matters in carbonaceous chondrites
\citep{iom}.
The molecule distribution in carbonaceous chondrites that includes 
sugars, deoxysugars, Soluble Organic Matter (SOM), and IOM may be inconsistent with the conventional building-up type synthesis \citep{isa_2021}.
However, the synthesis model proposed in our present paper could be consistent with the organic molecule distribution in carbonaceous chondrites.
Detailed comparison of the theoretical prediction 
with the distribution of organic matter species
(sugars, deoxysugars, amino acids, their derivatives, and IOM) 
is left for future study
\footnote{When the final products are discussed, the effect of
the activation energy need to be taken into account as discussed in Sect.~\ref{subsec:E_a}.}.

\section{Conclusions} \label{sec:conclusion}

In order to investigate photochemistry by
intermittent UV irradiation on the surface of icy particles in the protoplanetary disk,
we have developed a new Monte Carlo code to simulate reaction sequences of organic molecules,
using a graph-theoretic matrix model.
Our model does not assume a chemical reaction network in advance.
Instead, 1) we list up all the possible reactions at every step, 
2) select one of the reactions
with a weighting factor depending on the bond energy change and temperature,
assuming a warm environment where all the species can interact with one another without spatial separation,
3) and repeat this process to follow a reaction sequence.

We tested 26 patterns of starting molecules with 
the fiducial case of 7 CH$_2$O and 20 H$_2$O.  
For each pattern, we have performed $10^5$--$10^6$ Monte Carlo simulation runs consisting of
a reaction sequence of 70 UV phase steps and 30 post UV phase steps, 
considering icy dust particles that occasionally diffuse to an upper disk region
and are exposed to UV radiation from a host star (Figure~\ref{fig:disk}).

We focused on the analysis of the synthesis of 5-C sugars (ribose) and related molecules such as deoxysugars.
We have found that our simulation results would be consistent with
the past UV photochemistry experiments to synthesize ribose/deoxyribose
and the organic molecules in carbonaceous chondrites,
however, through a different pathway from 
the conventional step-by-step formose-type reactions that have been suggested by the past studies,
as illustrated in Fig.~\ref{mechanism}. 
Our results are summarized as follows.

\begin{enumerate}
\item Sugar abundance rapidly increases after UV irradiation is turned off.
Ribose (5-C sugars) and 4-C sugars show the similar evolution except for a slightly higher abundance for 4-C sugars.
The peak abundance values are not affected by UV photon energy ($E_{\rm UV}$) and the temperature ($T_{\rm p}$) at the post UV phase,
as long as $E_{\rm UV} \ga$ 1~eV (the equivalent temperature is $T_{\rm U} \ga 10^4$ K)
and $T_{\rm p} \la 10^3$~K.

\item During the UV phase, loosely-bonded O-rich large molecules are continuously created and destroyed.
In the post UV phase, these molecules are unstable and decomposed by
reactions cleaving O-O bonds and replacing C-OH by C-H, toward stable structures.
Molecule sizes become smaller with decreasing O/C.
Some fraction of the decaying molecules form sugars with $\rm O/C = 1$ and $\rm H/C = 2$.
We argue that the sugars should be stabilized by transformation to the cyclic form.

\item Our results would be consistent with
the past UV photochemistry experiments 
and carbonaceous chondrites.
However, the sugar synthesis pathway in our results 
contrasts with the previously suggested formose-type reactions where the molecular size
grows step by step and molecules at every step are stable at room temperature.

\item 
The synthesized ribose abundance does not depend on specific forms of starting molecules,
but is regulated mostly by H/O ratios of total starting molecules.
The abundance is peaked at $\rm H/O \simeq 2$,
such as a set of formaldehyde ($\mathrm{CH_2O}$) and water molecules ($\mathrm{H_2O}$). 

\item Ribose and deoxyribose are simultaneously synthesized with similar abundance, which is consistent with the past experimental results (deoxyribose is not produced via formose-type reactions).
Because the distribution generated by UV irradiation includes
divergent molecules and chemical structures of ribose and deoxyribose are similar, it is reasonable that their abundances are also similar.
\end{enumerate}

Our simulation suggests that most of the molecules finally become complex C-rich molecules, which
is similar to IOM, the major component of organics in carbonaceous chondrites.
The co-synthesis of ribose and deoxyribose we found may also be consistent with the chondrites. 
Detailed comparison with carbonaceous chondrites as well as with experiments is 
left for future study.

\begin{acknowledgements}
We thank the referee for helpful comments to improve the manuscript.
Because this research is a new challenge for us, we have had a lot of discussions with many colleagues.
We thank Hiroshi Naraoka, Yoshihiro Furukawa, Yasuhito Sekine, and Junko Isa
for comments from cosmochemistry, Ryuhei Nakamura for comments from chemistry, 
Yuri Aikawa, Hideko Nomura, and Kenji Furuya for comments from theoretical astrochemistry, 
Satoshi Yamamoto and Nami Sakai for comments from observations of interstellar organic molecules,
Hiroshi Kobayashi, Seiichiro Watanabe, Yuka Fujii, and Masahiro Ogihara for insightful theoretical discussions. 
We also thank Kimito Funatsu, Masashi Aono, Yoshi Oono, and Yu Komatsu for helpful advice
for developing the new scheme presented here.
This research is supported by JSPS Kakenhi 21H04512, MEXT Kakenhi 18H05438,
and MEXT “Program for Promoting Researches on the Supercomputer Fugaku”
(From clouds to stars and planets: toward a unified formation scenario).
\end{acknowledgements}


\end{document}